\documentclass[11pt]{article}

\usepackage{verbatim}
\usepackage{amsmath}

\usepackage{amssymb}					
\DeclareSymbolFontAlphabet{\amsmathbb}{AMSb}

\usepackage{simplewick} 
\usepackage{mathtools}

\usepackage{braket}

\usepackage{latexsym}
\usepackage{lscape} 
\usepackage{epsfig}
\usepackage{color}
\usepackage{graphicx}
\usepackage[table]{xcolor}
\usepackage{enumerate}
\usepackage{hhline}
\usepackage{subfig}
\usepackage{multirow}
\usepackage{mathrsfs}
\usepackage{dsfont}
\usepackage{longtable}

\usepackage{amscd}
\usepackage{cite}

\allowdisplaybreaks

\renewcommand{\arraystretch}{1.5} 

\newcommand{\red}[1]{\textcolor{red}{#1}}      
\newcommand{\blue}[1]{\textcolor{blue}{#1}}  

\definecolor{cadmiumgreen}{rgb}{0.0, 0.42, 0.24}

\definecolor{copper}{rgb}{0.72, 0.45, 0.2}

\definecolor{bleudefrance}{rgb}{0.19, 0.55, 0.91}

\definecolor{amethyst}{rgb}{0.44, 0.16, 0.39}

\definecolor{celestialblue}{rgb}{0.29, 0.59, 0.82}

\definecolor{cinnabar}{rgb}{0.89, 0.26, 0.2}

\definecolor{cyan}{rgb}{0.0, 0.72, 0.92}

\definecolor{dandelion}{rgb}{0.94, 0.88, 0.19}

\definecolor{darkcyan}{rgb}{0.0, 0.55, 0.55}

\definecolor{darkgreen}{rgb}{0.0, 0.2, 0.13}

\newcommand{\orange}[1]{\textcolor{orange}{#1}}

\DeclareMathSymbol{\mg}{\mathrel}{symbols}{"1D}

\newcommand{\ga}{\alpha}
\newcommand{\gb}{\beta}
\newcommand{\gd}{\delta}

\newcommand{\gl}{\lambda}

\newcommand{\gt}{\tau}
\newcommand{\go}{\omega}
\newcommand{\gz}{\zeta}

\newcommand{\gG}{\Gamma}
\newcommand{\gD}{\Delta}

\newcommand{\gL}{\Lambda}

\newcommand{\gO}{\Omega}

\newcommand{\Tr}{\mbox{Tr}}

\newcommand{\beq}{\begin{equation}}
\newcommand{\eeq}{\end{equation}}
\newcommand{\barr}{\begin{array}}
\newcommand{\earr}{\end{array}}
\newcommand{\equ}[1]{\begin{gather} #1 \end{gather}}

\newcounter{oldcounter}

\newcommand{\ba}[2]{\[\begin{array}{#2}\label{#1}}
\newcommand{\ea}{\end{array}\]}
\newcommand{\be}{\begin{equation}}
\newcommand{\ee}{\end{equation}}
\newcommand{\bea}{\begin{eqnarray}}
\newcommand{\eea}{\end{eqnarray}}

\usepackage{tikz}
\usetikzlibrary{calc}

\newcommand{\dd}{\text{d}}

\usepackage{tikz}

\makeatletter
\newcommand*\bigcdot{\mathpalette\bigcdot@{.5}}
\newcommand*\bigcdot@[2]{\mathbin{\vcenter{\hbox{\scalebox{#2}{$\m@th#1\bullet$}}}}}
\makeatother

\usepackage{hyperref} 
\usepackage{cancel}

\begin{document}

\thispagestyle{empty}

\phantom{}
\bigskip
\begin{center}
{\Huge 
\bf{
Abelian scalar theory\\[1.5ex] at large global charge
}
}
\\[6pt]\vspace{3cm}
{\large
{\bf{Orestis Loukas}\footnote{E-mail: orestis.loukas@cern.ch}
}
\bigskip}\\[0pt]
\vspace{0.23cm}
{\it 
Albert Einstein Center for Fundamental Physics\\
      Institute for Theoretical Physics\\
      University of Bern,\\
      Sidlerstrasse 5, \textsc{ch}-3012 Bern, Switzerland
}
\\[1ex]
\bigskip
\end{center}

\vspace{1cm}

\subsection*{\centering Abstract}

We elaborate on 
Abelian 
complex scalar 
models, which are dictated by natural actions (all couplings are of order one),
at fixed and large global $U(1)$ charge
in an arbitrary number of dimensions.
The ground state $\Ket{v}$ is coherently constructed by the zero modes
and the appearance of a centrifugal potential is quantum mechanically verified.
Using the path integral formulation we systematically analyze the quantum fluctuations around $\Ket{v}$
in order to derive an effective action for the Goldstone mode, which
becomes perturbatively meaningful 
when the charge is large.
In this regime we explicitly show that the whole construction is stable against quantum corrections,
in the sense that any higher derivative couplings to Goldstone's tree-level action are suppressed by appropriate powers of the large charge.

\newpage 
\setcounter{page}{1}
 \setcounter{footnote}{0}
\tableofcontents
\newpage

\section{Introduction}


In 
 \cite{Alvarez-Gaume:2016vff}
the authors argued that under certain (mild) assumptions 
a strongly coupled scalar theory can be found effectively at weak-coupling
in the sector of fixed and large global charge $Q_0$\,. 
This was motivated by  the previous observation  \cite{Hellerman:2015nra} that a perturbative expansion in the example of the three-dimensional $O(2)$ model at the conformal fixed point exhibits $1/Q_0$ as a good controlling parameter\footnote{In natural units, $Q_0$ is conveniently dimensionless.}, even when the original coupling is large.

\begin{figure}
\begin{center}
\includegraphics[width=11cm,height=11cm,keepaspectratio]{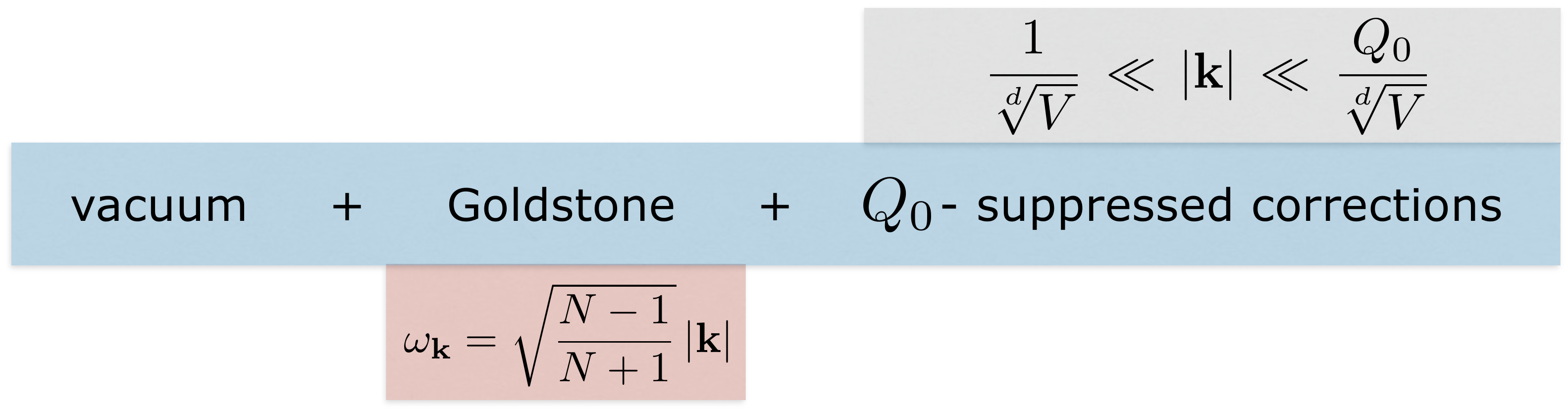}
\caption{A schematic diagram for  
the three ingredients entering the action of a complex scalar theory at fixed global charge $Q_0$\,.
When the hierarchy among the fundamental scales in the top right  corner applies ($V$ being the $d$-dimensional space volume), 
there is a stable condensated vacuum $\Ket{v}$ with a fluctuating Goldstone field $\chi(x)$ around it. This Goldstone mode exhibits a non-trivial dispersion relation $\go_\chi$ (depicted here  
for a pure $\vert \phi \vert^{2N}$ potential).
}\label{fig:FlowDiagramSUMMARY}
\end{center}
\end{figure}

In an arbitrary number of space-time dimensions $D=d+1$\,,   the idea was made precise and generalized by 
focusing on the class of $O(2n)$ vector models.
Fixing $\nu\leq n$ of the global charges implies then a spontaneous symmetry breaking
at a non-trivial vacuum state $\Ket{v}$\,, but also the appearance of Goldstone bosons 
with generalized dispersion relations
(see also the discussion in \cite{Nicolis:2011pv} and \cite{Watanabe:2013uya}).
Most interestingly, it was shown that 
by appropriately U$(\nu)$-rotating in field space, we can always arrange, such that 
the vacuum expectation value \textsc{(vev)} is assigned in one complex direction only, which in turn defines an Abelian subsector of the theory.
Only one parameter 
$\mu=\braket{\dot\psi}$ characterizes the time-evolution of
this \textsc{vev}.
At large $U(1)$ charge, 
this non-constant background dominates the infrared physics around $\Ket{v}$ via one relativistic Goldstone boson exhibiting a non-Lorentz invariant dispersion:
$\go_\chi (\textbf k ) =  c \,\vert \textbf k \vert$ with speed of light $c=c(\mu)<1$.
In Figure~\ref{fig:FlowDiagramSUMMARY} we present a schematic summary of  those  findings, where
 the contributions of the classical vacuum, the Goldstone and the higher derivative corrections to the effective action at fixed charge  are listed in decreasing order of $Q_0$\,.

In this paper we wish to investigate more closely the aforementioned 
 U(1) sector, which is predominantly present in scalar theories 
for any number of fields $n$ and 
fixed charges $\nu\leq n$\,.
In principle, we can start from any $O(2)$-symmetric (not necessarily scale invariant) smooth scalar potential $V(\vert\phi\vert)$\,, $\phi\in \mathbb C$\,. 
For concreteness we focus on polynomial potentials, 
\equ{
\label{NaturalPolynomialPotential}
V(\vert\phi\vert)=\sum_{i=1}^{N} \gl_{i} \vert \phi \vert^{2i}
\quad,\quad \gl_i \sim \mathcal O(1)
~,
}
because other classes of models, e.g.\ \textsc{dbi} actions including square roots, can be often (or especially at large charge) polynomially expanded.
The naturalness  condition on the coupling constants reminds us that these potentials are to be thought of as descending from some effective Wilsonian action with generic couplings dictated solely by the symmetries of the problem. 
In the present context the action associated to \eqref{NaturalPolynomialPotential} sets the initial high scale $\boldsymbol \gL$\,, from which we srtart to observe our model.
Since the 
behaviour of the $Q_0$-leading outcome at each stage of our derivation does not change,  we will additionally set $\gl_{i}=0$ for  $1<i<N$\,, in order to simplify further the calculations.
Our starting point is thus the  Hamiltonian for a complex scalar field in flat space,
\equ{
\label{O2Hamiltonian_FlatFieldSpace}
H = 
\int \dd^d x \, \mathcal H = \int \dd^d x \left( \vert\pi_\phi\vert^2 + \vert \nabla \phi \vert^2  + M^2 \, \vert \phi \vert^2 
+ \frac{2^N}{2N} \gl\, \vert \phi \vert^{2N} \right)
~,
}
commuting with the U$(1)$-charge 
\equ{
\label{U1ChargeDEFINITION}
Q =\int \dd^d x \, \rho =  -i \int \dd^d x \left(\pi \phi - \pi^*\phi^* \right)
\quad , \quad 
[H ,Q ] = 0
~.
}
For all the formal derivations we only need to assume that we are working in a finite space volume $V \sim \mathcal O(1)$ with vanishing boundary conditions \textsc{(bc)}\,. Hence, it makes sense to define the charge density as $\rho_0 = Q_0/V$\,.
As the Hamiltonian \eqref{O2Hamiltonian_FlatFieldSpace} enjoys in real degrees of freedom \textsc{(dof)} an $O(2)$ symmetry, we shall also refer to it as the $O(2)$ model.

Starting from such an action at scale $\boldsymbol \gL$, we can write down an effective action below 
that scale in a sector of fixed charge $Q=Q_0$, which at large $Q_0$ admits a fully perturbative treatment.
The existence of a perturbative expansion in a natural situation, as it is generically described by the potential \eqref{NaturalPolynomialPotential}, poses by itself a non-trivial statement.
Consequently, our objective is summarized as providing alternative derivations and further evidence in the quantum field theoretic \textsc{(qft)} context for the existence  and well-definedness of the large-charge vacuum of \eqref{O2Hamiltonian_FlatFieldSpace} as well as of the sensible $1/Q_0$ expansion around it.

\subsubsection*{Overview of the paper}


First, 
in Section \ref{sc:GroundStatePhysics}
it is shown that
imposing fixed charge as an operator identity in the zero-mode sector of the theory is equivalent to the anharmonic isotropic oscillator problem in quantum mechanics $(\textsc{qm})$ at fixed angular momentum on the Euclidean plane.
This fact manifests itself in the su(1,1) symmetry algebra of the vacuum state  $\Ket{v}$ at $T=0$\,.
That  observation enables us to use both flat as well as generalized coherent states to explicitly construct the large-charge vacuum $\Ket{v}$ via the zero modes, to begin with.
In particular, we derive for both cases  the appearance of a centrifugal barrier $\frac{Q_0^2}{2v^2}$ in the classical potential at fixed $Q_0$, which 
creates the non-trivial minimum to expand around.


Once the zero mode story at fixed charge is well understood, we proceed in Section \ref{sc:QuantumFluctuations} to discuss the quantum fluctuations on top of $\Ket{v}$\,.
These constitute the $Q_0$-subleading corrections to the zero-mode coherent state,
such that the full large-charge quantum state is constructed \cite{MurayamaLecture} as 
\equ{
\label{FullCoherentStateSCHEMA}
\Ket{Q_0} = 
\prod_{\vert \textbf k\vert > 0}^{\vert\boldsymbol \gL\vert} \Ket{\textbf k} \Ket{v}
~,
}
where each $\Ket{\textbf k}$ stands for an ordinary \textsc{qft} coherent state constructed on top of the condensated vacuum $\Ket{v}$\,.
For the purpose of discussing the quantum theory of those momentum fluctuations $\textbf k$, we follow a different path compared to the canonical quantization implemented in 
\cite{Alvarez-Gaume:2016vff} or the coset construction invoked in \cite{Monin:2016jmo}\,. 
We use the path integral formulation and realize that the fixed charge physics can be interpreted as the Fourier transform at level $Q_0$ of the grand-canonical ensemble at finite chemical potential $i\theta$\,.
As expected in the $U(1)$ sector, we find two modes, a massive field $a(x)$ and a massless $\chi(x)$ with an energy gap dictated by $Q_0$\,.
Subsequently, integrating out the massive mode we are able to write down a quantum action that is quadratic in the Goldstone field
and assert in the large charge regime that all higher derivative corrections are suppressed by certain powers of $1/Q_0$\,.

%
%

In Appendix \ref{appendix:QM} we provide some basic technical machinery for the computations we perform.
In Appendix \ref{Appendix:Compactifications} some elements of the compactification of large-charge theories are illustrated.
Apart from that, in Appendix \ref{appendix:FreeTheory} we discuss the free field theory at fixed charge using the language of holomorphic formalism.

\section{Ground state physics}
\label{sc:GroundStatePhysics}

In this section we shall analyze the ground state at large charge.
The process describing charge fixation and the associated symmetry breaking is termed Bose-Einstein condensation \textsc{(bec)}.
It can be treated classically by considering  Hamilton's equations of motion \textsc{(eom)} while incorporating the finite charge density as a canonical  variable.
Alternatively,
we  show here, that fixing the charge by the momentum zero modes implies a restriction of the initial Hilbert space of the theory.
Inside the restricted Hilbert space
we shall explicitly construct the state of large charge both as a flat and as a generalized coherent state in quantum mechanics \textsc{(qm)}. 
This construction is important, as it constitutes  the leading approximation to the actual coherent state of the fixed-charge problem, schematically given in Eq.\ \eqref{FullCoherentStateSCHEMA}\,.

\subsection{Zero-mode Hamiltonian}
\label{ssc:ZeroModeHamiltonian}

%
As far as the zero modes are considered 
we can choose to parametrize our scalar field in real components,
\equ{
\phi =\frac{1}{\sqrt 2} \left(\phi_R + i \phi_I\right)
\quad , \quad
\pi =\frac{1}{\sqrt 2} \left(\pi_R - i \pi_I\right)
~.
}
By subsequently going   to momentum space 
via
\equ{
\label{FourierTrafo_FieldMomenta}
\phi_i(t,\textbf x) =\,\, \sqrt{\frac{m}{V} } \sum_\textbf{k} \text{e}^{i \textbf k \cdot \textbf x} \, \tilde \phi_i(t,\textbf k)
\quad , \quad
\pi_i(t,\textbf x) =\,\, \sqrt{\frac{1}{m V} } \sum_\textbf{k} \text{e}^{i \textbf k \cdot \textbf x} \, \tilde \pi_i(t,\textbf k)
\quad \text{for} \quad 
i = R,I
~,
}
where $V$ denotes the total space volume and $m$ is a \textsc{qm} mass parameter, 
we can explicitly separate the zero mode piece in  the Hamiltonian \eqref{O2Hamiltonian_FlatFieldSpace}\,:
%
\equ{
 \frac{1}{2m} \left( \tilde\pi_R(t,\textbf 0)^2 + \tilde\pi_I(t,\textbf 0)^2  \right)
+ \frac{m M^2}{2}   \left(  \tilde\phi_R(t,\textbf 0)^2 + \tilde\phi_I(t,\textbf 0)^2 \right)
+
\frac{m^{N}}{V^{N-1}} \frac{\gl }{2N} \left( \tilde\phi_R(t,\textbf 0)^2 + \tilde\phi_I(t,\textbf 0)^2  \right)^N.
\label{OperatorQFT:Phi4:Hamiltonian_ZeroModesOnly}
}
In addition, to fix the U$(1)$ charge \eqref{U1ChargeDEFINITION} starting from eigenvalue equation
\equ{
\label{ChargeOperatorIdentityDEFINITION}
 Q \vert Q_0 \rangle = 
\sum_{\textbf k}  \left[ \tilde \phi_I(t,-\textbf k) \tilde \pi_R(t,\textbf k) - \tilde \phi_R(t,-\textbf k)\tilde \pi_I(t,\textbf k) \right]  \vert Q_0 \rangle 
= Q_0 \vert Q_0 \rangle
~,
}
we impose
\equ{
\label{ZeroModes:ChargeFixation}
\left[ \tilde\phi_I(t,\textbf 0)\tilde\pi_R(t,\textbf 0) - \tilde\phi_R(t,\textbf 0) \tilde\pi_I(t,\textbf 0) \right] \vert v \rangle 
 \overset{!}{=} Q_0 \vert v \rangle
~,
}
i.e.\ we require that our \textsc{qm} coherent state introduced in \eqref{FullCoherentStateSCHEMA} fulfils charge conservation as an exact operator identity.

In total,
we observe that, after  relating the position $x,y$ and conjugate momenta $p_x,p_y$ of an isotropic oscillator 
to the zero modes of the $O(2)$ model in Eq.\ \eqref{O2Hamiltonian_FlatFieldSpace}, according to 
\equ{
 x ~ \leftrightarrow ~\tilde\phi_R(t,\textbf 0) 
 \quad , \quad
 y ~ \leftrightarrow ~\tilde\phi_I(t,\textbf 0)
\quad \text{ and } \quad
 p_x ~ \leftrightarrow ~\tilde\pi_R(t,\textbf 0) 
 \quad , \quad
 p_y ~ \leftrightarrow ~\tilde\pi_I(t,\textbf 0)
 ~,
 \label{QM_QFT_FieldCorrespondance}
}
the problem of fixing the charge and constructing the coherent state 
leads precisely to the \textsc{qm} case of the anharmonic isotropic oscillator  in 2D at fixed planar angular momentum.
Thus, we turn to investigate this problem in greater detail.
In the following of the zero-mode discussion, we interchangeably use the concepts of U(1) charge and 2D angular momentum.

\subsection{Quantum mechanics at fixed angular momentum}
\label{ssc:CoherentCOnstructionOfZeroModeVacuum}

Following the identification scheme \eqref{QM_QFT_FieldCorrespondance}
we express the zero modes in oscillator space by a standard canonical transformation to the circular basis of left- and right-movers (for the conventions used see Eq.\ \eqref{QM:CircularOperatorBasis} in appendix).
In particular, charge conservation \eqref{ZeroModes:ChargeFixation}  in oscillator space reads
\equ{
\label{ZeroModes:ChargeFixation_OperatorIdentity}
Q  \Ket{v} =
\left( a^\dagger_L a_L  - a_R^\dagger a_R\right) \Ket{v}  \overset{!}{=} Q_0 \Ket{v}
~.
}
To construct 
 such a ground state $\Ket{v}$, we need to select from  the full set of
number operator eigenstates,
\equ{
N_{i} \Ket{n_L,n_R} = a_{i}^\dagger a_{i}\Ket{n_L,n_R} 
=
n_{i} \Ket{n_L,n_R} 
\quad \text{ with }
\quad n_i \in \mathbb N_0
\quad \text{ for } \quad
i=L,R
~,
}
which form the original Hilbert space $\mathfrak H$ of the isotropic oscillator,
only those satisfying
\equ{
Q \Ket{n_L,n_R} = Q_0 \Ket{n_L,n_R}
\quad \Rightarrow \quad
n_L - n_R = Q_0
~.
}
This subset defines the restricted Hilbert space  
\equ{
\label{RestrictedHilbertSpaceDEFINITION}
\mathfrak H_{Q_0} \, = \,\left\lbrace \Ket{n}_{Q_0} = \Ket{n_R+Q_0,n_R} \big\vert_{n_R=n}  ~ \forall n \in \mathbb N_0  \right\rbrace
~.
}

Consequently, the problem of finding the zero-mode vacuum of scalar $O(2)$\,, Eq.\ \eqref{O2Hamiltonian_FlatFieldSpace}\,, under (first class) constraint \eqref{ZeroModes:ChargeFixation}
reduces to variationally constructing the coherent state of  
the quantum anharmonic isotropic oscillator problem 
\equ{
\label{IsotropicOscillator:Hamiltonian}
H = \frac{\textbf p^2}{2m} + \frac{m M^2}{2} \textbf r^2 +  \frac{m^N}{V^{N-1}} \,\frac{\gl}{2N}\, \textbf r^{2N}
\quad , \quad
\textbf r = (x,y)~,~\textbf p = (p_x,p_y)
~,
}
inside the Hilbert space \eqref{RestrictedHilbertSpaceDEFINITION}\,. For the remainder of this section we shall  
denote the zero-mode Hamiltonian \eqref{OperatorQFT:Phi4:Hamiltonian_ZeroModesOnly} simply by $H$.

\subsubsection*{Introducing the su(1,1) generators}

The structure-preserving operators for the Hilbert space defined in Eq.\ \eqref{RestrictedHilbertSpaceDEFINITION}
are found  \cite{inomata1992path} to be
\equ{
\label{su11Generators_LeftRightBasis}
K_0 
= \frac12 \left( a_L^\dagger a_L + a_R^\dagger a_R \right) 
\quad , \quad
K_+ 
=
a_L^\dagger a_R^\dagger
\quad , \quad
K_- 
=
a_L a_R
~.
}
They satisfy the standard angular momentum algebra
\equ{
\label{SU11algebra}
[K_0,K_\pm]=\pm K_\pm
\quad , \quad
[K_-,K_+]=2K_0
~,
} 
and are thus identified with the generators of the  Lie algebra of su(1,1).
A relevant realization \cite{klauder1985applications} of this algebra for the isotropic (an)harmonic oscillator problem is given by
\equ{
\label{su11_DimensionfulRealization}
\begin{align}
K_0 =&\,\, \frac14 \left( m\go \,\textbf r^2 + \frac{1}{m\go} \textbf p^2 \right) ~,&
\\[1ex]
K_+ =&\,\, \frac14 \left( -m\go\, \textbf r^2 + \frac{1}{m\go} \textbf p^2 + i\, (\textbf r \cdot \textbf p + \textbf p \cdot \textbf r) \right) 
\nonumber
~,~
& K_- = &\,\, \frac14 \left( -m\go \,\textbf r^2 + \frac{1}{m\go} \textbf p^2  - i\, (\textbf r \cdot \textbf p + \textbf p \cdot \textbf r) \right)
~,
\nonumber
\end{align}
}
One can verify that the three operators in this representation 
satisfy the algebra of Eq.\ \eqref{SU11algebra}\,, once the canonical commutation relations among $\textbf r, \textbf p$ are assumed.
The frequency $\go$\,, which is needed to match the dimensions properly\footnote{Conveniently, in natural units the su(1,1) generators are dimensionless, $[K_0]=[K_\pm]=0$\,.}, 
can be thought of as a Bogoliubov parameter in the sense outlined in Appendix \ref{ap:CanonicalTrafos}\,.
As we will shortly derive, this parameter drops from all physical quantities. 
Since we are interested in the physics inside $\mathfrak h_{Q_0}$\,, it is sensible to express the
initial Hamilton operator \eqref{IsotropicOscillator:Hamiltonian} only in terms of the structure-preserving operators \eqref{su11Generators_LeftRightBasis}\,. Using 
representation \eqref{su11_DimensionfulRealization} 
this leads to the following ``radial'' expression for 
the Hamiltonian under investigation: 
\equ{
\label{Hamiltonian_su11Basis}
H =  
\go \left( 1+\frac{M^2}{\go^2} \right) K_0 + \frac\go2 \left( 1 - \frac{M^2}{\go^2} \right) \left( K_+ + K_- \right)
+ \frac{1}{V^{N-1} } \frac{\gl}{2N}\, \frac{1}{\go^N} \left(2K_0 - K_+ - K_- \right)^N
~.
}

At this stage we invoke yet another realization of the su(1,1) algebra, which is particularly well-adjusted for our purposes, first  used by  \cite{MLODINOW1980314,MLODINOW1982387} in a similar context of large-$N$ expansion,
namely the so-called Holstein-Primakoff representation,
\equ{
\label{HolsteinPrimakoffREPRESENTATION}
K_0 = a^\dagger a + \frac{Q_0}{2}
\quad, \quad
K_+ = a^\dagger \sqrt{Q_0+a^\dagger a}
\quad, \quad
K_- = \sqrt{Q_0+a^\dagger a}\, \,a
~,
} 
where $a,a^\dagger$ are canonical ladder operators satisfying $[a,a^\dagger]=1$\,. 
It is intuitive\footnote{In fact, one can rigorously show \cite{arecchi1972atomic} that at $Q_0 \rightarrow\infty$ the angular momentum algebra can be contracted (by taking formal limits) to the harmonic oscillator algebra.
In particular, this implies a contraction of the corresponding radial coherent states defined in \eqref{RadialCoherentStateDEFINITION} to the Glauber state \eqref{GlauberStatesDEFINITION} via $\lim_{Q_0\rightarrow\infty}\sqrt{Q_0}\, \gz \, = \ga$\,.}
to think that the auxiliary mathematical space upon which $a,a^\dagger$ act is identified (to leading order)
with the restricted Hilbert space $\mathfrak h_{Q_0}$ introduced in \eqref{RestrictedHilbertSpaceDEFINITION}\,:
\equ{
 N_R \equiv a^\dagger a \equiv N \quad,\quad
 \text{in particular} \quad
a\Ket{0}_{Q_0} = 0
~.
}
After this realization
we are  in a position to construct the vacuum state for the zero modes as a flat (Glauber) coherent state:
\equ{
\label{GlauberStatesDEFINITION}
\Ket{\ga} \, := \, \text{e}^{-\frac12\vert\ga\vert^2}\,\text{e}^{\ga\,a^\dagger }\, \vert 0 \rangle_{Q_0}
\quad \text{ with } \quad
a \Ket{\ga} = \ga  \Ket{\ga}
\quad , \quad \ga \in \mathbb C
~.
}
Here, $\Ket{0}_{Q_0}$ signifies the vacuum of the restricted Hilbert space $\mathfrak H_{Q_0}$ defined in Eq.\ \eqref{RestrictedHilbertSpaceDEFINITION}\,.
The associated coherent energy, i.e.\ the \textsc{vev} of Hamiltonian \eqref{Hamiltonian_su11Basis} w.r.t.\ $\Ket{\ga}$\,, reads
\equ{
\label{FlatCoherentEnergy_DEFINITION}
\begin{align}
R(\ga,\go) \equiv
\braket{H} := \frac{\braket{\ga \vert H \vert\ga} }{ \braket{\ga \vert \ga} }
=  &\,
\omega \left( 1+\frac{M^2}{\go^2} \right) \left(\ga^2  + \frac{Q_0}{2} \right) +
 \omega \left( 1-\frac{M^2}{\go^2} \right) \ga\,  \sqrt{Q_0 + \ga^2} 
\\
&
+
\frac{1}{V^{N-1} \omega^N}  \frac{\gl}{2N} \left( 2\ga^2  - 2\ga \sqrt{ Q_0+\ga^2}   + Q_0\right)^{N}
~.
\nonumber
\end{align}
}
Relying on the underlying U(1) symmetry we may always choose $\ga \in \mathbb R$\,.
Expressing then the coherent parameter $\ga$ in terms of the radial amplitude $v$ and the Bogoliubov frequency $\go$,
\equ{
v^2 = \frac{m}{V} \langle \textbf r^2 \rangle 
\quad \longleftrightarrow \quad
\ga = \frac{Q_0 - V\go\, v^2}{2\sqrt{V\go} \,v }
~,
}
we can rewrite \eqref{FlatCoherentEnergy_DEFINITION} as 
\equ{
\label{QM:ClassicalRuthian}
\mathcal R(v,\rho_0) \equiv 
\frac{R(v,\rho_0)}{V} = 
\frac{\rho_0^2}{2v^2} + \frac{M^2}{2} v^2 + \frac{\gl}{2N} v^{2N}
\quad , \quad 
\rho_0 = \frac{Q_0}{V}
~.
}
Therefore, we see that with the present coherent construction the centrifugal potential, familiar from classical mechanics, 
is precisely reproduced by fixing the charge.
Notice also, that the final result written in terms of the field theoretic quantities $\rho_0$ and $v$ is independent of the Bogoliubov parameter $\go$, as it should be.


Alternatively, we could directly proceed from Eq.\ \eqref{Hamiltonian_su11Basis} by constructing generalized (or ``radial'') coherent states of Perelomov-type compatible with the algebra \eqref{SU11algebra}
following the prescription first presented in~\cite{perelomov1972,PhysRevA.38.191}: 
\equ{
\label{RadialCoherentStateDEFINITION}
\vert \gz \rangle\, := \, (1-\vert \gz \vert^2)^\frac{Q_0}{2}\, \text{e}^{\gz K_+} \vert 0 \rangle_{Q_0}
~,
}
where the coherent $\gz = -\tanh(\theta/2)\, \text{e}^{i\varphi}$ can be parametrized in terms of  the SU(1,1) group manifold parameters $\theta\in\mathbb R$\,,\,$\varphi\in[0,2\pi]$\,.
Using (again we arrange for $v\in\mathbb R$)
\equ{
v^2 = \frac{m}{V} \braket{\textbf r^2}  = \frac{m}{V}  \frac{\braket{\gz \vert \textbf r^2 \vert \gz} }{\braket{\gz \vert \gz}} 
\quad \longleftrightarrow \quad
\cosh\theta = \frac{Q_0}{2V\go\, v^2} + \frac{V\go\,v^2}{2Q_0}
~,
}
we recognize that the generalized \textsc{vev} of the quadratic kinetic term, 
\equ{
\braket{\frac{\textbf{p}^2}{2m} } = \frac{ \braket{\gz\vert\frac{\textbf{p}^2}{2m} \vert\gz} }{\braket{\gz\vert\gz}}
=
V \frac{\rho_0^2}{2v^2 }
~,
\label{RadialCoherent:BarrierTerm}
}
precisely reduces to the classical centrifugal barrier. 
Note at this point that we did not need to rely on the large charge assumption to derive the appearance of this barrier term in the classical potential $\mathcal R$, only at its finiteness.
On the other hand, as explained in \cite{gerry1983large} and outlined in Appendix \ref{app:QM_Identities} we find for the original potential $\textbf r^{2N}$ the relation
\equ{
\label{su11MatrixElement_potential}
\braket{\textbf r^{2N} } = \braket{\textbf r^2}^N + \sum_{j=1}^{N-1}\mathcal O \left(Q_0^{N-j}\right)
~,
}
so that in total the generalized coherent \textsc{vev} $ \frac{ \braket{\gz\vert H \vert\gz} }{\braket{\gz\vert\gz}}$
agrees \textit{to leading order}\footnote{To higher orders in $Q_0$
the Perelomov-type coherent  energy differs from the previous flat one.
This is due to the non-trivial coherent measure associated to $\Ket{\gz}$\,.
Since we aim to formulate the path integral in \textsc{qft} context with flat coherent states, we do not pursue the radial coherent construction in this work further.}
with the centrifugal potential derived in Eq.\  \eqref{QM:ClassicalRuthian} via the Holstein-Primakoff construction.

Eventually minimizing the classical centrifugal potential derived  in \eqref{QM:ClassicalRuthian} we fix the value of the radial amplitude $v$ or equivalently of the coherent parameters $\ga\,,\,\gz$ in \eqref{GlauberStatesDEFINITION}\,,\,\eqref{RadialCoherentStateDEFINITION} respectively:
\equ{
\label{PhiN:BEC_minimum}
\begin{align}
\frac{ \partial \mathcal R(v,\rho_0)  }{\partial v} \overset{!}{=}0 
\quad \Rightarrow \quad
v^2 \, =\, \left(\frac{\rho_0^2}{\gl} \right)^{\frac{1}{N+1} } \left[1 - \frac{M^2}{\rho_0^2}\,v^4 \right]^\frac{1}{N+1} 
~.
\end{align}
}
Therefore, we see that the coherent ground state $\Ket{v}$ of fixed and large charge 
naturally constructed by the zero modes to first approximation, is uniquely determined  by the centrifugal potential in Eq. \eqref{QM:ClassicalRuthian}\,.
Its coherent energy \eqref{FlatCoherentEnergy_DEFINITION} at the variational minimum has by itself an $1/Q_0$-expansion for non-vanishing mass parameter $M$\,: 
\equ{
\label{ClassicalRuthian_Q0Expansion}
\mathcal R(v,\rho_0)
=
\frac{N+1}{2N} \left( \gl\, \rho_0^{\,2N} \right)^\frac{1}{N+1} \, + \, \frac{M^2}{2} \, \left(\frac{\rho_0^{\,2} }{\gl} \right)^\frac{1}{N+1}
-
\frac{M^4}{4(N+1)}  \left(\rho_0^{\,-2N+4} \, \gl^{-2N+1}  \right)^\frac{1}{N+1}
+
\mathcal O \left( \rho_0^{\, \frac{-4N+6}{N+1} }  \right)
.
}
For $M=0$ all the classical expressions at the coherent vacuum become exactly solvable.


In total, we learn from the study of the zero modes that just by restricting ourselves to the Hilbert space of fixed $Q_0$ naturally leads to the emergence of su(1,1) symmetry.
Exploiting this symmetry it is possible to construct explicitly the classical vacuum $\Ket{v}$ of our theory (at any finite value of the fixed charge)
as a coherent state, whose energy includes the crucial centrifugal barrier term \eqref{RadialCoherent:BarrierTerm}.
This ensures the existence of a non-trivial minimum (Eq.\ \eqref{PhiN:BEC_minimum}), around which we are going to find a stable perturbative expansion at large charge.

\section{Quantum fluctuations}
\label{sc:QuantumFluctuations}

In this section we analyze the  quantum  fluctuations $\Ket{\textbf k}$ in Eq.\ \eqref{FullCoherentStateSCHEMA},
on top of the  large-charge vacuum $\Ket{v}$ constructed in the previous section.
For this purpose we revert to the path integral formalism and generically define the fixed charge partition sum by 
\equ{
\label{FixedChargePartitionSumDEFINITION}
Z_{Q_0} (\gb) := \Tr\, \gd \left( Q- Q_0 \right) \, \text{e}^{-\gb H} 
=
\int \frac{\dd\theta}{2\pi} \, \text{e}^{-i\theta Q_0} \, \Tr \,\text{e}^{i\theta Q} \,\text{e}^{-\gb H}
\equiv 
\int \frac{\dd\theta}{2\pi} \, \text{e}^{-i\theta Q_0} \, Z^\theta(\gb)
~,
}
with $H$ being the Hamiltonian of the initial system and $Q$ the  operator associated to the charge we wish to fix.
To make contact with thermal field theory literature \cite{kapusta2006finite} as well as the holomorphic treatment in Appendix \ref{appendix:FreeTheory}, we perform a standard Wick rotation, $\int \dd^D x\, \rightarrow \int_0^\gb \dd\gt \int\dd \textbf x$\,,
while identifying $\gb=T^{-1}$\,.
$Z^\theta(\gb)$ stands for the partition sum of the grand-canonical ensemble at finite chemical potential\footnote{This imaginary chemical potential is formally conjugate to the charge operator $Q$\,.} $i\theta$\,.

In the following, we focus on our paradigm of the $O(2)$ model in Eq.\ \eqref{O2Hamiltonian_FlatFieldSpace} with Lagrangian density 
\equ{
\label{O2LagrangianDensity}
\mathcal L[\phi,\phi^*] = \partial_\mu\phi^* \partial^\mu \phi - M^2 \vert\phi\vert^2 - \frac{2^N}{2N}\,\gl\, \vert\phi\vert^{2N}
~.
}
We shall find that the quantum theory at fixed $Q_0$ has one massless (Goldstone) mode and a massive field.
Once we integrate out the massive mode, an effective action quadratic in the massless field emerges.
We will make sure that all higher derivative corrections to this action are suppressed in the large charge regime.

\subsection{The ground state in the semi-classical treatment}

Before proceeding with the quantum analysis though, we want to relate the semi-classical formulation to the previous \textsc{qm} derivation of the zero-mode ground state. 
In order to fully exploit the U(1) symmetry it is most convenient, if we revert to the polar field basis
\equ{
\label{RadianAnsatzFieldSpace}
\phi(x) 
= \tfrac{1}{\sqrt 2} \left( \phi_R(x) + \phi_I(x) \right)
= \tfrac{1}{\sqrt 2}\, r(x) \,\text{e}^{i\psi(x)}  
\quad , \quad 
\psi \in [0,2\pi]
~.
}
Note that this basis change in field space is related \cite{Argyres:2009em} to a non-trivial Jacobian factor in the path integral: 
\equ{
\label{PathIntegralJacobianDEFINITION}
 \mathcal D \phi_R\, \mathcal D \phi_I = 
 \mathcal D r \, \mathcal D \psi \, \exp \left\lbrace \int \frac{\text d^Dx}{\gb V} \log \left(\sqrt{\frac{V}{\gb}} \, r(x) \right) \right\rbrace
~.
 %
}
Similar to  ordinary polar coordinate transformation, this path integral Jacobian depends only on the radial field.
In this polar basis the U(1) charge density corresponding to \eqref{U1ChargeDEFINITION} becomes
\equ{
\label{EuclideanChargeDensity}
 \rho = - \,i \left( \phi^* \partial_0 \phi  - \phi \partial_0 \phi^*\right) = r^2 \dot \psi
 ~,
}
so that
the fixed charge condition in Eq.\ \eqref{ZeroModes:ChargeFixation} at $T=0$ takes the classical form of angular momentum conservation,
\equ{
\label{VEVconfiguration}
\braket{\rho}  = \braket{r^2 \dot\psi} = v^2 \braket{\dot\psi} \overset{!}{=} \rho_0
\quad \Rightarrow \quad
\mu \equiv \braket{\dot\psi} = \frac{\rho_0}{v^2}
~,
}
where \textsc{vevs} are taken w.r.t.\ the large-charge vacuum $\Ket{v}$ constructed in Section \ref{sc:GroundStatePhysics}\,.
Therefore, we conclude that finite charge density $\rho_0\neq 0$ 
in field space
means not only a non-vanishing radial \textsc{vev} $\braket{r^2}=v^2$ associated to the spontaneous symmetry breaking, but it also implies the existence of a non-constant background for the angular field\footnote{As in the case of constant  \textsc{vev}\,, we can use the U(1) symmetry to set an overall integration constant $\psi_0$ to zero.}:
\equ{
\label{AngularVEV}
\psi = \mu \, t + ...
~.
}

This time-dependent background  is already manifest at the grand-canonical level, which is used as an intermediate step in the transformation prescription \eqref{FixedChargePartitionSumDEFINITION} to obtain the canonical physics at fixed charge.
The grand-canonical Lagrangian density $\mathcal L^\theta(\gb)$ associated  to \eqref{O2LagrangianDensity} 
is given by \cite{kapusta2006finite} 
\equ{
\label{PhiN:GrandCanonical_Lagrangian}
\mathcal L^\theta [r, \psi ] =
\frac12 \left[ \partial^\mu r\, \partial_\mu r + r^2\, \partial^\mu \psi \,\partial_\mu \psi\, \,+\, \,2i\, \frac{\theta}{\gb}\, r^2 \,\partial_0 \psi - \left(\frac{\theta^2}{\gb^2} + m^2 \right) r^2 - \frac{\gl}{N}\, r^{2N} \right]
~.
}
It is crucial to realize that the appearance of an effective mass $\frac{\theta^2}{\gb^2}$ in the radial direction acts as a ``natural'' regulator for the Fourier integral in the defining Eq.\ \eqref{FixedChargePartitionSumDEFINITION}\,. 
Implementing  \textsc{vev}-configuration \eqref{VEVconfiguration} 
we next calculate the 
canonical potential at zero temperature in the sector of finite $Q_0$\,:
\equ{
\label{PhiN:CanonicalPotentialDEFINITION}
\mathcal V(v,\mu) =
- \frac{1}{\gb V}\,\log Z_{Q_0} (\gb)
\big\vert_{\gb\rightarrow\infty}
=
-\frac{v^2}{2}\, \mu^2 + \left(\frac{M^2}{2}\, v^2 + \frac{\gl}{2N}\, v^{2N} \right) 
~,
}
where in the canonical partition function,
\equ{
Z_{Q_0} ( T\text{=}0 ) = \int \frac{\dd\theta}{2\pi} \, \text{e}^{-i\theta Q_0} \left(v\, \text{e}^{\gb V \mathcal L^\theta[v,\mu]} \right)
~,
}
the path integral Jacobian \eqref{PathIntegralJacobianDEFINITION} was evaluated at the ground-state to $J=v$. In fact, in the present condensate arrangement this factor stems from the familiar Jacobian appearing in  ordinary polar-integration in 2D. 

Legendre-transforming \eqref{PhiN:CanonicalPotentialDEFINITION} we end up again with the classical centrifugal potential\footnote{In classical mechanics such a function is called a Ruthian, hence the letter $\mathcal R$\,. Among the three classical variables $\lbrace v,\mu,\rho_0\rbrace$ related via constraint Eq.\ \eqref{VEVconfiguration}, the Lagrangian formalism treats $(v,\mu)$ as dynamical, whereas the Hamilton formulation uses $(v,\rho_0)$\,.} of Eq.\ \eqref{QM:ClassicalRuthian},
\equ{
\mathcal R (v,\rho_0) = 
\rho_0 \, 
\mu
 \,+\, \mathcal V(v, \mu )
=
\frac{\rho_0^2}{2v^2} + \frac{M^2}{2}\, v^2 + \frac{\gl}{2N}\, v^{2N}
~,
}
whose minimum has been determined in \eqref{PhiN:BEC_minimum}\,.
At this minimum, charge conservation \eqref{VEVconfiguration} would then imply for the angular \textsc{vev}:
\equ{
\label{PhiN:BEC_minimumMu}
\mu =  \langle \dot\psi\rangle = \frac{\rho_0}{v^2}
=
\left( \gl \, \rho_0^{N-1} \right)^{\frac{1}{N+1} } 
\left[ 1 + \frac{M^2}{N+1} \left(\frac{1}{\gl\,\rho_0^{\,N-1} }\right)^\frac{2}{N+1} + ...
\right]
~.
}
All in all, we see that the formal definition \eqref{FixedChargePartitionSumDEFINITION} reproduces indeed at $T=0$ the coherent state result obtained by the \textsc{qm} treatment of Section \ref{ssc:CoherentCOnstructionOfZeroModeVacuum}.

\subsection{The full quantum action}

Henceforth, we set $M=0$, i.e.\ we 
consider a pure $\vert\phi\vert^{2N}$ potential, just for computational simplicity. 
For discussing the fluctuations (i.e.\ momentum non-zero modes)  in the real fields 
it also suffices to first focus on $\theta=0$\,,
and afterwards incorporate the
effect of $\theta$-integration in Eq.\ \eqref{FixedChargePartitionSumDEFINITION} in the form of a quantum correction.

Starting from the radial ansatz \eqref{RadianAnsatzFieldSpace} we write the fluctuations on top of $\Ket{v}$ in the form of
\equ{
\phi(x) = \tfrac{v+\ga(x)}{\sqrt 2} \, \text{e}^{\,i\left[\mu \,t\, +\chi(x)/v\right]}
~,
}
such that the grand-canonical Lagrangian  $\mathcal L^\theta$ at $\theta=0$ breaks into powers of fluctuating $\ga(x)$ and $\chi(x)$\,, schematically:
\equ{
\label{OrderPiecesGrandCanonical}
\mathcal L^{\theta=0}[\phi,\phi^*] = - U(v,\mu) + \sum_{i=1}^{2N}  \mathcal L^{(i)} [v,\mu\,;\ga,\chi] 
~.
}
Because of the U(1) symmetry being realized as a shift symmetry in the $\chi$-direction, there can be no potential for the $\chi$ field.
Hence, this field is going to play the role of the Goldstone boson in the effective action we aim at.

Generally, linear terms $\mathcal L^{(1)}[v,\mu\,;\ga,\chi] $  in the fluctuations are to be anticipated, since we are not expanding around the grand-canonical vacuum (i.e.\ the minimum of $U(v,\mu)$ in expansion \eqref{OrderPiecesGrandCanonical}).
However, all those terms drop from the action.
In detail, by the zero-mode assumption any term linear in $\ga$  vanishes:
\equ{
\int \text{d}^D x\, \mathcal L^{(1)}[v,\mu\,; \ga] \, \sim \, \int \text{d}^D\, x\, 
\sum_{k_\mu \neq 0} \text{e}^{-ik\cdot x} \ga(k)
 \, \sim \,  
\sum_{k_\mu \neq 0} \delta^D(k)\,  \ga(k) =0
~,
}
while the appropriate periodic or vanishing \textsc{bc} for the $\chi$ field in volume $V=L^d$,
\equ{
\chi(\gt+\gb,\textbf x) = \chi (\gt,\textbf x) \quad \text{ and }\quad
\chi(\gt, \vert \textbf x \vert \rightarrow L ) = 0~,
}
implies the vanishing of $\mathcal L^{(1)}[v,\mu\,;\chi]$ as a total derivative (remember that due to U$(1)$ symmetry only powers of $\partial\chi$ are allowed).
Analyzing next the  Lagrangian piece quadratic in $\ga(x),\chi(x)$\,,
\equ{
\label{QuadraticLagrangian}
\mathcal L^{(2)} [v,\mu\,; \ga,\chi] = 
\frac12\partial^{\mu} \ga \,\partial_{\mu} \ga - \frac{1}{2} \left[ \left(2N-1\right) \gl v^{2N-2} - \mu^2\right] \ga^2
+
\frac12\,\partial^{\mu} \chi \, \partial_{\mu} \chi \,+\, 2 \mu\, \ga\, \dot\chi 
~,
}
we observe that $\ga(x)$ is a massive field with a quadratic potential (written with the help of Eq.\  \eqref{PhiN:BEC_minimum} and \eqref{PhiN:BEC_minimumMu} in terms of $\gl$ and $\rho_0$)\,,
\equ{
\mathcal V^{(2)} [\ga,\chi] :=
\left(N-1\right)\left( \gl \rho^{\,N-1}_0 \right)^\frac{2}{N+1} \,\ga^2\, 
- \,
2 \left( \gl \rho^{\,N-1}_0 \right)^\frac{1}{N+1} \, (\partial_0 \chi) \, \ga
~.
} 
To fully appreciate the significance of this radial mode we need to move to the minimum of $\mathcal V^{(2)}$\,,
%
\equ{
\label{MasslessPhiN_alphaMinimum}
\ga_\text{min} = 
\frac{1}{(N-1)} \,
\left(\frac{1}{ \gl \rho^{\,N-1}_0 } \right)^\frac{1}{N+1}
\left(\partial_0 \chi \right)
~,
} 
in order to
re-expand $\mathcal L^{(i)}[v,\mu\,;\ga,\chi]$ around 
$a \equiv \ga-\ga_\text{min}$\,.
For the remainder of this section we will treat $\chi(x)$ as a background field.
It is convenient to relabel the various pieces $\mathcal L^{(i)}[v,\mu;\ga,\chi]$ for $i=2,...,2N$ of the grand-canonical Lagrangian in Eq.\ \eqref{OrderPiecesGrandCanonical} 
according to powers of $a$ alone: 
%
%
\begin{align}
\label{PhiN:StationaryAction_aOrderPiecesLagrangian}
%
\mathcal L_\chi^{(0)}[a]= &\,
\blue{
\frac12 \left[ \left( \frac{N+1}{N-1} \right) \left(\partial_0 \chi\right)^2 - \left(\nabla\chi\right)^2 \right] 
}\,+
\red{
 \left(\frac{1}{\gl\, \rho^{\,2N}_0 } \right)^\frac{1}{2N+2}
\left[ \frac{N+1}{3(N-1)^2}\left(\partial_0\chi\right)^3 - \frac{1}{N-1}\left(\partial_0\chi\right)\left(\nabla\chi\right)^2\right]
} 
\nonumber
\\
&\,\,
+\, 
{
\frac{1}{2(N-1)^2}\, 
\left(\frac{1}{ \gl \, \rho_0^{\,N-1} }\right)^\frac{2}{N+1} 
\partial_\mu (\partial_0\chi) \partial^\mu (\partial_0\chi)}
\,+\, \mathcal O\left(\left(\frac{1}{\rho_0}\right)^{\frac{2N}{N+1} }\right)
\nonumber
\\[1ex]
\mathcal L_\chi^{(1)}[a] = &\,
-
a\,
 \left(\frac{\gl}{\rho^{\,2}_0 } \right)^\frac{1}{2N+2}
\left[  \left(\frac{N-2}{N-1} \right) \left(\partial_0\chi\right)^2 + (\nabla\chi)^2 \right]
%
{ -
\frac{1}{(N-1)}
 \left(\frac{1}{\gl\, \rho^{\,N-1}_0 } \right)^\frac{1}{N+1}
a \left( \partial_\mu \partial^\mu\partial_0\chi \right) 
}
\,+ 
\,\mathcal O\left( \frac{1}{\rho_0} \right) 
\nonumber
\\[1ex]
\mathcal L_\chi^{(2)}[a] = &\,
\frac{1}{2} \partial_\mu a \,\partial^\mu a -
\frac{a^2}{2} \,\bigg\lbrace 2 \left(N-1\right) 
 \left( \gl \,\rho_0^{\,N-1} \right)^\frac{2}{N+1}
 \orange{
\,+\,4 
\left( \gl^\frac{3}{2} \, \rho_0^{N-2} \right)^\frac{1}{N+1}\,
 \left(\partial_0\chi\right) \,\,+
 }
\nonumber
\\
&\,\, \orange{
+ 
\left( \frac{\gl}{\rho_0^2} \right)^\frac{1}{N+1}
\, \left[\frac{4N^2-9N+4}{N-1} \left( \partial_0\chi \right)^2 +\left(\nabla\chi\right)^2 
 \right]
}
\,+\, 
\mathcal O\left( \left(\frac{1}{\rho_0} \right)^{\frac{N+2}{N+1} }\right) 
\bigg\rbrace
\nonumber
\\[1ex]
%
\mathcal L_\chi^{(3)}[a] = &\,\,
 \frac{a^3}{3} \left[ -
\left( \gl^\frac{5}{2}\, \rho_0^{2N-3} \right)^\frac{1}{N+1} 
 \left( 2N^2-3N+1 \right) 
 {+ 
\left( \gl^2\, \rho_0^{\,N-3} \right)^\frac{1}{N+1}
 \left( 4N^2-8N+3 \right)(\partial_0\chi)
 } \,+\, 
 \mathcal O \left( \left(\frac{1}{\rho_0}\right)^\frac{3}{N+1} \right)  \right]
\nonumber
\\[-1.5ex]
\,\vdots\,&
\nonumber
\\[-1.5ex]
\mathcal L_\chi^{(m)}[a] = &\,\,
-a^m \left[ 
\mathcal O \left(\rho_0^{\frac{2N-m}{N+1} }\right)
\, + \, 
\mathcal O \left(\rho_0^{\frac{N-m}{N+1} }\right) (\partial_0\chi)
 \,+ \, 
 \mathcal O \left( \rho_0^{-\frac{m}{N+1} } \right) 
 (\partial_0\chi)^2 + ...\right]
\nonumber
\\[-2ex]
\vdots\,&
\nonumber
\\[-2ex]
\mathcal L_\chi^{(2N)}[a] = &\,\,
- \frac{\gl}{2N} a^{2N}
~.
\end{align}
The colours help to keep track of the high scale origin $\boldsymbol \gL$ (i.e.\ before integrating out the $a$ field) of the first few terms in the final expression \eqref{PhiN:EffectiveActionChiRESULT} for the $\chi$-effective action.
In total, we clearly recognize that $a(x)$ becomes a very massive field ($m_a \sim \rho_0$) in the large charge regime with a
time-independent tree-level propagator\footnote{There are slightly different ways to define the propagator of the $a$ field.
In particular, one can either consider a time-dependent mass by including background dependent corrections to \eqref{PhinN_aTreeLevelPropagator} or start by solving the inhomogeneous Klein-Gordon equation deduced from \eqref{QuadraticLagrangian}\,.
Of course, the final result is insensitive to any given choice.} 
\equ{
\label{PhinN_aTreeLevelPropagator}
D^{-1}_0(k)_a \,= 
-k_0^2+\textbf k^2 + 2 \left(N-1 \right)  \left( \gl \,\rho_0^{\,N-1} \right)^\frac{2}{N+1}  
\,\equiv\,
-k_0^2 + \omega_a(\textbf k)^2
~,
}
and higher-loops on a non-constant, $\partial\chi$-dependent, background.
Therefore, it does not influence the low-energy dynamics around $\Ket{v}$\,, and hence
it makes sense to integrate out this massive mode. 
In other words, we have to (perturbatively) solve the path integral of massive $a^{2N}$-theory in presence of ``external'' current,
\equ{
\label{ChiCurrent}
j_{\chi}(x) = 
-
\left(\frac{\gl}{\rho_0^{\,2}}\right)^{\frac{1}{2N+2} } 
\left[  \left(\frac{N-2}{N-1} \right) \left(\partial_0\chi\right)^2 + (\nabla\chi)^2 \right] +...
~.
}
Note that the $\chi$-current is $\rho_0$-suppressed, only because we have shifted to the minimum of $\ga$ fluctuations determined by \eqref{MasslessPhiN_alphaMinimum}.

\subsubsection*{Integrating out the massive mode}

At this stage, our goal is to fully integrate out the $\rho_0$-massive field $a(x)$ in order to write down an effective action for the light mode $\chi(x)$ fluctuating around $\Ket{v}$ at scales below $\boldsymbol \gL$\,. This $\chi$-effective action would 
look like
\equ{
\label{EffectiveActionChi_DEFINITION}
\begin{align}
S_\text{eff}\left[\chi\right]_{Q_0}\, := &\,\,
-\gb V \mathcal V(v, \mu) + S[\chi]
+\log Z_0 + 
\left\langle  S_I [a,\chi] + S_{\theta-\text{eff} } [a,\chi] + S_\text{polar}[a,\chi] 
\right\rangle_a
~, 
\end{align}
} 
where
the canonical classical potential $\mathcal V$ was computed in \eqref{PhiN:CanonicalPotentialDEFINITION},
$S[\chi]$ is the $a(x)$-independent part of the action in \eqref{PhiN:StationaryAction_aOrderPiecesLagrangian}
and
 $Z_0$ denotes the tree-level partition sum of the $a(x)$ field, 
\equ{
\label{HarmonicPartitionSum_aMode}
Z_0\left[ j_\chi; \chi \right] 
 =  \int \mathcal Da \,\, \text{e}^{S_0[a,\chi\,;\, j_\chi] }
 \quad, \quad
 S_0[a,\chi\,;\, j_\chi] = \int \dd^D x \left(\mathcal L_\chi^{(1)}[a]+\mathcal L_\chi^{(2)}[a] \right)
~.
}
Expectation values are taken in this context w.r.t.\ the harmonic action $S_0$\,: 
\equ{
\left\langle  ...\right\rangle_a 
:=\, \frac{ \int \mathcal Da \,\, \text{e}^{S_0[a,\chi;j_{\chi}]}\,\, ...}{  \int \mathcal Da \,\, \text{e}^{S_0[a,\chi;j_{\chi}]} }
~~.
}
$S_I$ includes the $a(x)$-loops described by $\mathcal L^{(i>2)}_\chi[a]$ and 
for ease of reference
we have also summarized the effect of $\theta$-integration \eqref{FixedChargePartitionSumDEFINITION} and polar Jacobian \eqref{PathIntegralJacobianDEFINITION} in $S_{\theta-\text{eff} }$ and $S_\text{polar}$\,, respectively.

%

Performing first the Gaussian path integral in $Z_0$\,, we immediately find 
\equ{
\label{HarmonicPartitionSum_MassiveIntegratedOut}
\frac{\log  Z_0}{V} =
- \int
\frac{\text{d}^d\textbf k}{(2\pi)^d} \left[ \frac{\gb\, \go_a(\textbf k)}{2} 
+ \log\left(1-\text{e}^{-\gb \go_a(\textbf k) } \right) \right]
{+ 
\tfrac12 \int \dd^D x \int \dd^D y\,  j_{\chi}(x)\, D_0(x-y)_a j_{\chi}(y)
}
~.
} 
Since  $\go_a(\textbf{k}) \sim \rho_0^\frac{N-1}{N+1}$ (see Eq.\ \eqref{PhinN_aTreeLevelPropagator}), 
the $ \log\left(1-\text{e}^{-\gb \go_a(\textbf{k}) } \right)$-part is exponentially suppressed
and 
hence negligible. 
The vacuum energy of $a$-mode (first term) on the other hand is of tructable order, but irrelevant for the $\chi$-effective action. 
Furthermore, there is a non-vanishing source term, relevant for the $\chi$-effective action, which is expressed 
in terms of the position space propagator $D_0(x-y)_a$ for the $a(x)$ field.
To estimate its order we need the $\rho_0$-scaling of space-time propagator given in \eqref{SpaceTimePropagatorSCALING} as well as the scaling of $j_{\chi}(x)$ defined in \eqref{ChiCurrent}\,,
so that we find its contribution to the $\chi$-effective action being less than $\mathcal O \left( \rho_0^{-(2N+1)/(2N+2)} \right)$\,.

\subsubsection*{Momentum cut-off and suppression of $a$-loops}

At this point we turn to the interaction terms encoded in $S_I[a,\chi]$ of formula \eqref{EffectiveActionChi_DEFINITION}\,.
To be able to consistently integrate out the $a(x)$ field, we need to ensure that all pure-$a$ loops are suppressed, once $Q_0\gg1$\,.
%
The basic ingredients needed for a detailed computation of $a(x)^m$-loops  on the non-constant $\chi$-background are listed in appendix \ref{sc:ThermalSums}\,.
Here, we only note an outline of basic facts to establish the desired suppression.
Inspecting the various powers of $a^m$ in \eqref{PhiN:StationaryAction_aOrderPiecesLagrangian} at $\theta=0$\,, one explicitly records a decreasing scaling behaviour with increasing $m \geq 2$\,:
\equ{
\mathcal L^{(m)} \,\sim\, v^{2N-m}\, a^m \, + \, \text{ subleading }\,\partial_0\chi \text{ terms}
\quad,\quad
v \sim \mathcal O\left(\rho_0^{2/(N+1)} \right)
~.
}
Since the leading background-dependent quadratic term $\sim\int_X \left\langle a^2(x) \,\partial_0\chi(x) \right\rangle_a$ is kinematically vanishing, the remaining dominant loop is the pure-$a$ quartic interaction,
\equ{
\label{PhiN:LeadingQuarticInteraction}
\braket{ S_I^{(4)}[a] }_a \,\sim\,
v^{2N-4} \, \braket{\int \dd^Dx\, a^4(x)}
\sim \mathcal O \left(\left(\frac{\vert\boldsymbol\gL\vert^d}{\rho_0^\frac{1}{N+1}}\right)^2\right)
~,
}
where the
base Matsubara sum is resolved in Eq.\ \eqref{BasicLoop_MatsubaraSum_SCALING} of Appendix \ref{sc:ThermalSums}\,.
For the  integral over spatial momentum we need to introduce a cut-off, $\textbf k \leq \boldsymbol \gL$\,.
The scaling of \eqref{PhiN:LeadingQuarticInteraction} helps us then set the fundamental momentum cut-off at
\equ{
\label{TreeLevelMomentumCutOff_FINAL}
\vert\boldsymbol\gL\vert \, \sim\, \mathcal O \left(\rho_0^\gamma\right) ~: \quad
0\,\leq\, \gamma \,<\, \frac{1}{d\left(N+1 \right)}
\quad,\quad\text{fundamental cut-off}
~.
}
Notice here, that last inequality needs to be strict (in order for loop-suppression to be achieved), whereas the first inequality could well be saturated, 
as it merely arises from requiring a sufficiently large momentum cut-off $\boldsymbol \gL$ for interesting physics to appear around larger $Q_0$\,.

For the various loops of $a(x)$ we note that each (un)-contracted field combination brings at most (if not kinematically constrained)
\equ{
\label{PhiN:PaircontractionsScaling}
\braket{aa}_a
\, < \,
\mathcal O \left( \left(\frac{1}{\rho_0}\right)^\frac{N-2}{N+1} \right)
\quad , \quad
\braket{a}_a
\, < \,
\mathcal O \left( \left(\frac{1}{\rho_0}\right)^{\frac{2N-1}{2N+2}+\frac{1}{2d(N+1)}} \right)
~.
}
The derivation of the latter scaling is based on Eq.\ \eqref{LinearCurrentLoop_SCALING}\,.
Both unsaturated upper bounds\footnote{In this context, inequalities among different orders imply a $\rho_0$-dictated hierarchy.} are estimated at the cut-off \eqref{TreeLevelMomentumCutOff_FINAL}\,.
In addition,
due to 
the non-vanishing source $j_\chi(x)$ a finite number of corrections arises in each contraction, by taking namely  (multiple) functional derivatives of $\exp \left\lbrace \tfrac12 \int \dd^d x \int \dd^d y  j_\chi(x) D_0(x-y)_a j_\chi(y)\right\rbrace$\,. 
These terms however
are always quite subleading to the fully contracted term (the one surviving, when the current vanishes)
due to the $\rho_0$-suppressed current $j_\chi(x)$ defined in Eq.\ \eqref{ChiCurrent}\,.

\subsubsection*{Effect of Fourier integration}

To fully flow  to the canonical ensemble at fixed charge we need  to Fourier integrate as indicated in formal definition \eqref{FixedChargePartitionSumDEFINITION}\,.
From Eq.\ \eqref{PhiN:GrandCanonical_Lagrangian} we immediately recognize that this $\theta$-integral has a dominant Gaussian kernel, namely
\equ{
 Z_{Q_0}(\gb) 
=
\text{e}^{-\gb V \mathcal V(v,\mu)} \, \int \frac{\dd\theta}{2\pi}\, 
\text{e}^{- \frac{V}{\gb} \frac{v^2}{2}\, \theta^2 }\, 
\int \mathcal D a \mathcal D \chi\, \exp \left\lbrace \int \dd^Dx \sum_{i=2}^{2N} \mathcal L^{(i)}[a,\chi] \right\rbrace
~.
}
Due to the $\rho_0$-scaling of $v$ in Eq.\ \eqref{PhiN:BEC_minimum}\,, this results into   
physical quantities being exponentially suppressed already at $\theta \sim \mathcal O(1)$\,.

Indeed, after performing the $\theta$-integral, we find that 
the leading contribution is a pure $a$-term, 
$\rho_0^{(2N-4)/(N+1)}\braket{\left(\int \dd^Dx \, a^2(x)\right)^2}_a< \mathcal O(1)$\,,
which remains suppressed 
due to the cut-off \eqref{TreeLevelMomentumCutOff_FINAL}\,.
The first non-trivial correction to the $\chi$-effective action from this Fourier part amounts to 
%
\equ{
\braket{S_{\theta-\text{eff} }[a,\chi]}_a \equiv 
 \frac{-2}{\gb V}
 \left(\gl^2 \, \rho_0^{N-3} \right)^\frac{1}{N+1}
\left\langle
 \int_X a(x)^2 \int_Y \left(\partial_0\chi(y)\right) a(y) 
\right\rangle_a 
<
\mathcal O \left( \left(\frac{1}{\rho_0}\right)^{\frac{2N+1}{2N+2}+\frac{1}{2d(N+1)} }  \right)
~,
}
which lies beyond  the order  of the few first terms we have chosen to display  in Eq.\ \eqref{PhiN:EffectiveActionChiRESULT}\,.

\subsubsection*{Non-trivial path integral Jacobian}

Since we have expanded $\phi$ in polar field variables $(\ga,\chi)$\,, we need to account for  a non-trivial Jacobian factor in those fluctuations.
Effectively, it leads to an additional piece in the grand-canonical action \eqref{PhiN:GrandCanonical_Lagrangian} of the form stated in \eqref{PathIntegralJacobianDEFINITION}\,, which in the present setup with $r = v + \ga$ becomes 
\equ{
\text e^{\,S_\text{polar}[v,\ga]} =
v \,\exp \left\lbrace \int \frac{\text d^D x}{\gb V} \log \left(1 + \frac{\ga(x)}{v} \right) \right\rbrace 
~.
}
Notice that this  depends now on the $\chi$ field, because we have shifted $\ga(x)$ around \eqref{MasslessPhiN_alphaMinimum}\,.
Expanding the logarithm for large $v$ we calculate
the contraction of the pure-$a$ leading term by using \eqref{PhiN:PaircontractionsScaling},
\equ{
\frac{1}{v^2}\,
\int \frac{\dd^Dx}{\gb V}\left\langle a^2(x) \right\rangle_a
\, \sim \, 
\mathcal O \left( \frac{\vert\boldsymbol\gL\vert^d}{\rho_0
} \right)
\,<\, \mathcal O (1)
~,
}
proving consistent with integrating out the massive mode.
The relevant  terms for the $\chi$-effective action, 
$\braket{S_\text{polar,eff}[a,\chi]}_a \,< \,\mathcal{O}\left(\rho_0^{-2N/(N+1)} \right)$\,
start way beyond the order displayed  in Eq.\ \eqref{PhiN:EffectiveActionChiRESULT}\,.

\pagebreak

\subsection{Goldstone effective action}

Assembling all pieces outlined in the previous paragraphs we are now in the position to write down an effective action for the $\chi$ field alone.
This is thought of as the light mode entering the low-energy action around the large-charge vacuum $\Ket{v}$\,, obtained once we fully integrate out the massive mode.
As we shall conclude, it is possible to
analytically follow the derivation from the full quantum action at scale $\boldsymbol \gL$ presented in  \eqref{PhiN:StationaryAction_aOrderPiecesLagrangian} to the low-energy effective action around $\Ket{v}$\,, only due to the large charge assumption.
To easily keep track of this procedure we use colours that refer to the full action at scale $\boldsymbol \gL$. They show the field content a certain term from action  \eqref{PhiN:StationaryAction_aOrderPiecesLagrangian} brings down to the low-energy effective action, once the $a$-mode is integrated-out.
The first few terms in the $\chi$-effective action are summarized as

\equ{
\label{PhiN:EffectiveActionChiRESULT}
\begin{align}
S_\text{eff}\left[\chi\right]_{Q_0}\, =&\,\,
{
\gb V \, \frac{N-1}{2N} \, \left( \gl \, \rho_0^{\,2N} \right)^\frac{1}{N+1}
~+~    
}
\blue{
\int \dd^Dx\,
\bigg\lbrace\,\, \frac12 \left[ \left( \frac{N+1}{N-1} \right) \left(\partial_0 \chi\right)^2 - \left(\nabla\chi\right)^2 \right] 
}
\\[1.5ex]
&\,\,
\red{ +\,
\left(\frac{1}{\rho_0}\right)^{\frac{N}{N+1} } \left(\frac{1}{\gl}\right)^{\frac{1}{2N+2} }
\left[ \frac{N+1}{3(N-1)^2}\left(\partial_0\chi\right)^3 - \frac{1}{N-1}\left(\partial_0\chi\right)\left(\nabla\chi\right)^2\right]
} 
\nonumber
\\[1.5ex]
&\,\,
\orange{ 
-\frac{1}{4}
\left( \frac{\gG\left(\frac d2\right)}{(2\pi)^\frac{d}{2}}\,
 \frac{1}{d\sqrt{2\left(N-1\right)} } \right)
\frac{\vert\boldsymbol\gL\vert^d}{ \rho_0 }
 \int \dd^Dx \left[\frac{4N^2-9N+4}{N-1} \left( \partial_0\chi \right)^2 +\left(\nabla\chi\right)^2 
 \right]
 ~+\, ...
}
\nonumber
\end{align}
}

The low-energy action starts thus,  with the classical potential at fixed $Q_0$\,, defined in Eq.\ \eqref{PhiN:CanonicalPotentialDEFINITION}, which is provided here for the massless case.
Since in ordinary \textsc{qft} considerations we have chosen to work in Lagrangian formalism, note that this potential is \textit{not} equal to the centrifugal potential \eqref{PhiN:CanonicalPotentialDEFINITION}.
Rather, it is naturally identified as the leading vacuum contribution of the running Goldstone corresponding to the \textsc{vev} configuration \eqref{AngularVEV}.

Because of the global  U(1) symmetry, $\chi \rightarrow \chi + \xi$\,,
no mass-term can be generated perturbatively.
Therefore, we see that indeed $\chi$ plays the role of a Goldstone mode fluctuating around the vacuum $\Ket{v}$\,.
Its tree-level dispersion relation (blue term) manifestly breaks  Lorentz invariance.
As it is extensively commented in \cite{Alvarez-Gaume:2016vff,Nicolis:2013sga}\,, this breakdown of Lorentz covariance is to be fundamentally expected due to the rapidly rotating \textsc{vev} we had to assign in Eq.\ \eqref{AngularVEV} to the angular field, only in the temporal direction, in order to fulfil charge conservation condition \eqref{VEVconfiguration} at the classical level.
All derivative corrections to leading Goldstone's dispersion relation are also expected to break Lorentz invariance in the same manner.
On the other hand, the
spatial kinetic term is not influenced by this \textsc{vev} configuration, which means that in contrast to large spin theory, homogeneity in space is preserved to all perturbative orders (for recent developments in higher spin theories, also in the context of $O(2n)$ models, see e.g.\ \cite{Alday:2016jfr}).

The first correction to the leading Goldstone dispersion (red term)  comes from $\mathcal L^{(0)}_\chi[a] \equiv \mathcal L[\chi]$ in \eqref{PhiN:StationaryAction_aOrderPiecesLagrangian}\,;
already just by freezing the massive mode around the minimum \eqref{MasslessPhiN_alphaMinimum} (or by roughly substituting for $\ga(x)$ the solution to its classical \textsc{eom}).
For $N=2$ there are further corrections to the leading dispersion coming from $\mathcal L[\chi]$\,.
Otherwise, for $N\geq3$\,,
the next-to-leading correction is posed by the background dependent $a$-mass in $\mathcal L_\chi^{(2)}[a]$ of \eqref{PhiN:StationaryAction_aOrderPiecesLagrangian} (orange term)\,.
All in all, we see that for all those higher corrections the appearance of the charge density $\rho_0$ in the denominator gurantees the desired suppression, when the compatible momentum cut-off \eqref{TreeLevelMomentumCutOff_FINAL} is imposed.
The $\rho_0$-order at which further contributions from $S_I[a,\chi]$ as well as $S_{\theta,\text{eff}}[a,\chi]$ and $S_\text{polar,eff}[a,\chi]$ 
appear in the $\chi$-effective action, once the massive mode is integrated out, is sumamrized in Table~\ref{tb:Rho0OrderChiEffective}.

\begin{table}
\begin{center}
  \renewcommand{\arraystretch}{1.4}
 \begin{tabular}{|c|c|l|}
 \hline
\multicolumn{3}{|c|}{scaling of terms in $S_\text{eff}[\chi]_{Q_0}$}
\\ \hline
& high-scale term & description
\\ \hline\hline
&
\multirow{2}{*}{$\mathcal L^{(0)}_\chi$} & Goldstone dispersion relation 
\\ \cline{3-3}
& & tree-level correction(s) at $a=\ga_\text{min}$
\\ \cline{2-3}
$\rho_0$ & $\mathcal L^{(2)}_\chi$ &  background dependent correction to $m_a$
\\ \cline{2-3}
$\downarrow$ 
& $S_{\theta,\text{eff}}$ &  effect of Fourier integration
\\ \cline{2-3}
& $\mathcal L^{(1)}_\chi$ &  source term for linearly coupled $j_\chi(x)$ to $a$-field
\\ \cline{2-3}
& $\mathcal L^{(m\geq 3)}_\chi$ &  higher order loops
\\ \cline{2-3}
& $S_\text{polar,eff}$ &  effect of non-trivial Jacobian
\\ \hline
\end{tabular}
\renewcommand{\arraystretch}{1}
\end{center}
\caption{
In this table we schematically list (in decreasing strength of $\rho_0$) the order at which  contributions from the initial action at scale $\boldsymbol \gL$  appear in the low-energy $\chi$-effective action \eqref{EffectiveActionChi_DEFINITION}, once the $a$-mode is fully integrated out.
In particular, $\mathcal L^{(i)}_\chi$ refers to the $\chi$-dependent Lagrangian pieces in \eqref{PhiN:StationaryAction_aOrderPiecesLagrangian}, which contribute to $S_I[a,\chi]$\,.
The scaling of source term stemming from $\mathcal L^{(1)}_\chi$ is given below Eq.\ \eqref{HarmonicPartitionSum_MassiveIntegratedOut}\,.}
\label{tb:Rho0OrderChiEffective}
\end{table}



\section{Conclusions and Outlook}

To summarize we find that despite starting from a very excited state $\Ket{v}$ of large charge $Q_0$ (the condensate energy \eqref{ClassicalRuthian_Q0Expansion} scales with $Q_0$), it is still possible to write down a meaningful low-energy effective action around it.
{
Note at this point that in order to derive the form of the coherent state $\Ket{v}$ in Section \ref{sc:GroundStatePhysics} we only had to fix the charge $Q_0$ and not necessarily to assume that it is large.
On the other hand, the results presented in this section are crucially based on the large charge assumption to maintain a perturbative character.
}

\begin{figure}[t!]
\begin{center}
\includegraphics[width=9cm,height=9cm,keepaspectratio]{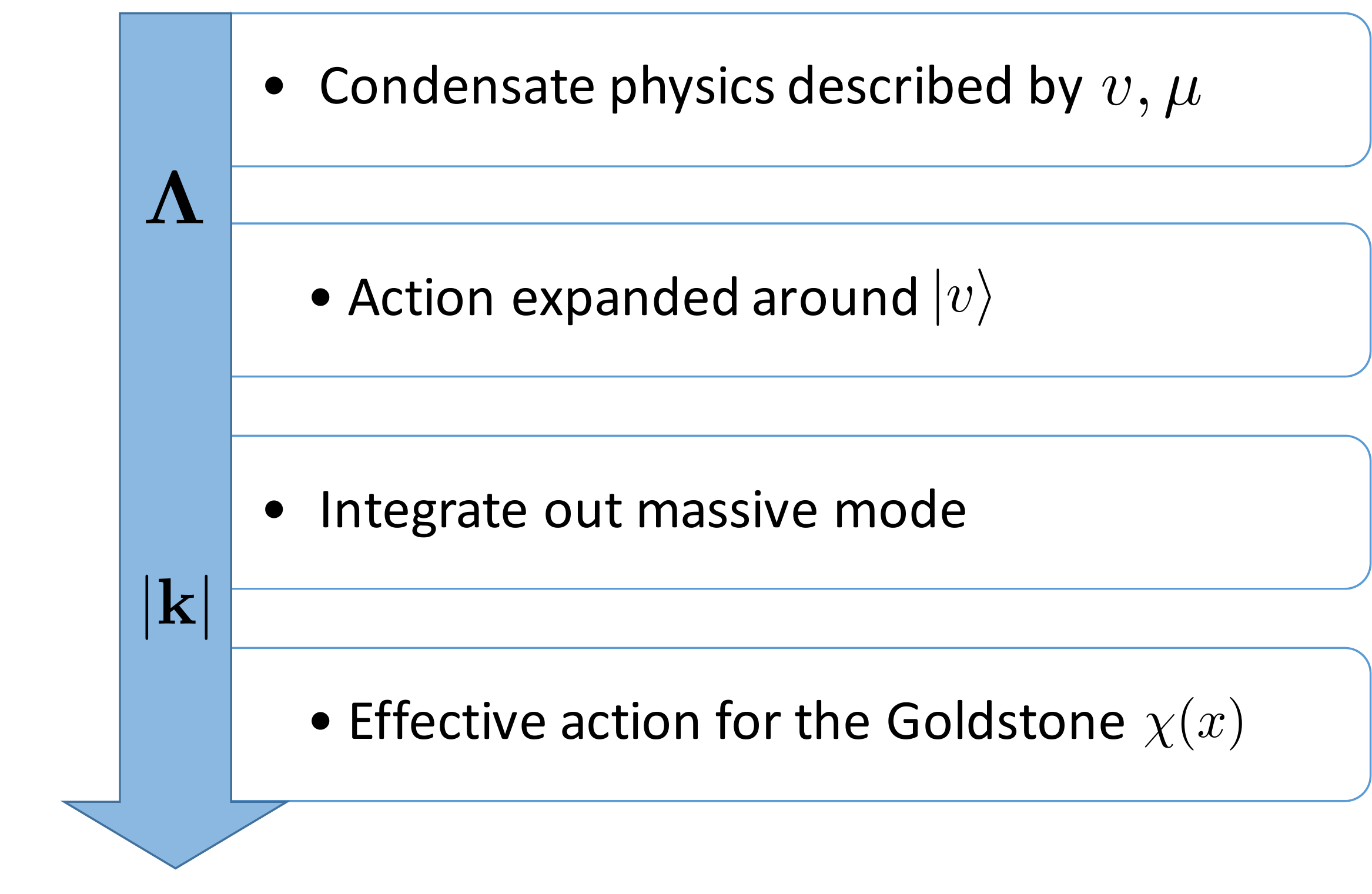}
\caption{A flow diagram depicting the steps followed  in Sections \ref{sc:GroundStatePhysics} and \ref{sc:QuantumFluctuations} to derive the low-energy effective action \eqref{PhiN:EffectiveActionChiRESULT} for the Goldstone mode, starting from the action \eqref{O2LagrangianDensity} expanded around \textsc{vev} configuration \eqref{VEVconfiguration} and integrating out the massive mode as dictated by Eq.\ \eqref{EffectiveActionChi_DEFINITION}. The order at which the diagram flows from top to bottom coincides with the scale of the various effects from higher to lower energies. 
}\label{fig:FlowDiagramConclusions}
\end{center}
\end{figure}

Approaching the low-energy regime around $\Ket{v}$ is achieved by fully integrating out one of the two physical \textsc{dofs}, the massive mode $a(x)$\,,
i.e.\ solving perturbatively the corresponding path integral $\int \mathcal D a$\,,
as instructed in Eq.\ \eqref{EffectiveActionChi_DEFINITION}\,.
There thus remains a light mode, the Goldstone $\chi(x)$, to describe the small-momentum\footnote{Here, ``small'' is taken in the sense of the momentum cut-off \eqref{TreeLevelMomentumCutOff_FINAL}.} excitations around $\Ket{v}$\,, encoded by $\Ket{ \textbf k}$ in Eq.\ \eqref{FullCoherentStateSCHEMA}\,.
The resulting low-energy effective action \eqref{PhiN:EffectiveActionChiRESULT} has a perturbative expansion in higher derivative couplings of the Goldstone field $\chi(x)$, which is controlled by the large charge $Q_0$\,.
In that sense, we verify that the construction is indeed
stable against quantum corrections 
under the hierarchy presented in 
Figure~\ref{fig:FlowDiagramSUMMARY} among the fundamental scales of the original problem.
Finally, in Figure \ref{fig:FlowDiagramConclusions} we diagrammatically summarize the steps we have followed to setup our large-charge perturbative expansion.  

%
After we have seen that at large charge there is a sensible perturbative expansion of our scalar theory in flat space, 
we can directly apply this in computing the vacuum energy of non-trivial space configurations,
i.e.\ when we compactify $\gd\leq d$ space dimensions on some manifold $\mathcal M$\,.
Apart from phenomenological purposes, the energy computation in such a topologically non-trivial setup becomes meaningful in particular, either when the theory is at a conformal fixed point
or  in the context of the Casimir force.


The first case has been already considered in Chapter 5 of \cite{Alvarez-Gaume:2016vff}, so we only briefly review it here.
The aim is to compute the anomalous dimension of the field operator in the three-dimensional $O(2)$ model
at the Wilson and Fischer conformal fixed point.
As it is argued there this amounts to calculating the vacuum energy of the vector model compactified on the two-sphere $S^2(R)$
with a potential $V=\frac{1}{4R^2}\vert \phi \vert^2+\tfrac{4}{3}\gl \, \vert \phi \vert^6$\,.
The main point to be emphasized is that the qualitative considerations leading to low-energy action \eqref{PhiN:EffectiveActionChiRESULT} are to be applied also, once the compactification space is $S^2(R)$ with the only major adjustment (this is justified in Appendix \ref{Appendix:SphereQuantization}, where the role of the free field is played to leading order after the condensate by the Goldstone mode $\chi(x)$) that the momentum is now discretized according to $\textbf k^2 ~\rightarrow ~ {l(l+1)}/{R^2}$\,.
Notice also that the appearance of an effective mass ${1}/{4R^2}$  because of the conformal coupling to non-vanishing curvature has only a subleading effect in every step of the preceding large charge derivation, as it falls into the general class of polynomial potentials in Eq.\  \eqref{NaturalPolynomialPotential}\,.


Regarding the Casimir effect the sensible energy to examine in a  
finite volume at finite temperature is the (thermodynamic) free energy\footnote{The upperscript $(p)$ is to remind us that quantities are to be computed under (spatial) periodic identifications.}
 $\mathcal F^{(p)}_V$ of the topologically non-trivial theory, after subtracting the topologically trivial (i.e.\ with vanishing spatial \textsc{bc}) free energy $\mathcal F^\text{van}_V$ inside the same volume $V$:
\equ{
\label{CasimirEnergyDEFINITION}
\mathcal F^{(p)}_V : =
-\left(\tfrac{1}{\gb} \, \log Z_{Q_0}(\gb) \bigg\vert_V^{(p)} \,
-
\tfrac{1}{\gb} \, \log Z_{Q_0}(\gb) \bigg\vert^\text{van}_V  \right)
~.
}
Generically, 
the crucial point in this situation is that our condensates $v$ and $\mu$ %
are not influenced by 
space identifications.
Concretely, despite having a non-constant $\psi$-background in \eqref{AngularVEV}, $\langle\psi\rangle = \mu \, t$\,, this evolves only in the time direction.
On the other hand, any meaningful compatification  extends at most in $d$ spatial dimensions.
Therefore, from the perspective of $\langle \psi\rangle$ the effect of the compactifiaction %
is not felt, because the leading condensate contribution eventually drops from defining equation \eqref{CasimirEnergyDEFINITION}\,.
As a relevant application a toy model for circle compactification is discussed in Appendix \ref{appedix:ToyModelS1}\,.

\subsubsection*{Outlook}

In this work the principles and many properties of scalar \textsc{qft} at large global charge were explored.
Together with the recent interest in the literature in that and related topics (see the references cited in the introduction), it becomes clear that there is a fruitful path to be followed in order to deeper comprehend more theories at large global charge(s). 
After having analyzed the scalar vector model, the first (obvious) generalization are matrix models with fields in the adjoint representation of the global symmetry group.

A further task we aim at addressing in a future work is to  fix the charge in a purely fermionic theory.
Similar to the  spirit that was presented in this work  for the scalar theory, we wish to understand whether and how we can obtain a perturbative expansion in a purely fermionic theory, when starting from some natural action with coupling constants of order unity.
The holomorphic formalism discussed in Appendix \ref{appendix:FreeTheory}\,, which is the natural formulation for fermionic \textsc{qft}, might turn out to be beneficial in this context.
We plan to come back to those open subjects, also as part of the extended collaboration of \cite{Alvarez-Gaume:2016vff}, in the near future.

\section*{Acknowledgments}
 The author would like to thank Simeon Hellerman, Mikko Laine and Uwe-Jens Wiese for valuable discussion and
 comments.
Furthermore, the author thanks Luis~Alvarez-Gaume, Domenico Orlando and  Susanne Reffert for collaboration on a related project as well as proof-reading the manuscript.

The work of O.L. is supported by the Swiss National Science Foundation (\textsc{snf}) under grant number \textsc{pp}00\textsc{p}2\_157571/1.

\appendix
\renewcommand\thesection{\Alph{section}}

\section{Important relations and technical remarks}
\label{appendix:QM}

\subsection{Canonical transformations}
\label{ap:CanonicalTrafos}

As customary in quantum mechanical manipulations we think of writing the Hamiltonian in Eq.\ \eqref{IsotropicOscillator:Hamiltonian}
in terms of ladder operators via canonical transformations to 
oscillator space,
\equ{
\label{QM:OscillatorOperatorBasis}
\begin{align}
a_x = \sqrt{\frac{1}{2m\go} } \left( m\go\, x + i p_x \right)
 \quad &, \quad
 a_x^\dagger = \sqrt{\frac{1}{2m\go} } \left( m\go\, x - i p_x \right) 
 ~,
\end{align}
}
and similarly for $y$-coordinate. From there we define the oscillator basis for circular polarization as
\equ{
\label{QM:CircularOperatorBasis}
\begin{align}
a_L = \tfrac{1}{\sqrt 2} \left( a_x - i a_y \right)
 ~ , ~
a_L^\dagger = \tfrac{1}{\sqrt 2} \left( a_x^\dagger + i  a_y^\dagger \right) 
 \quad , \quad
a_R = \tfrac{1}{\sqrt 2} \left( a_x + i a_y \right)
 ~ , ~
 a_L^\dagger = \tfrac{1}{\sqrt 2} \left( a_x^\dagger - i  a_y^\dagger \right) 
~.
\end{align}
}

\subsubsection*{Families of unitary equivalent Hamiltonians}

For the su(1,1) realization of anharmonic problem in Section \ref{ssc:CoherentCOnstructionOfZeroModeVacuum} we need to
remind ourselves that our initial Hamiltonian can always be canonically deformed.
Algebraically, this is realized \cite{Xiao:1198089} via performing a canonical transformation to rescale $\textbf r\,,\textbf p$ expanded as in \eqref{QM:OscillatorOperatorBasis}.
The new set $b=\lbrace b_x,b_y \rbrace$\,,\,$b^\dagger=\lbrace b_x^\dagger,b_y^\dagger \rbrace$ of creation and annihilation operators is obtained from the old one $a,a^\dagger$ through a \textit{Bogoliubov} transformation \cite{PhysRevD.29.643} for each $i=x,y$\,, 
charachterized by a continuous parameter $\vert t\vert <1$\,:
\equ{
\label{BogoliubovTrafoDEFINITION}
b_i = \frac{a_i-t\,a_i^\dagger}{\sqrt{1-t^2} }
\quad , \quad
b_i^\dagger = \frac{a_i^\dagger-t\,a_i}{\sqrt{1-t^2} }
~.
}
Setting $\gO := \frac{1-t}{1+t}$
the Bogoliubov transformation means that $\textbf r\,,\textbf p$ are rescaled as
\equ{
x = \frac{1}{\sqrt{2}} \left( a_x + a_x^\dagger \right) = \frac{1}{\sqrt{2 \gO } } \left( b_x + b_x^\dagger \right)
\quad , \quad
p_x =  \frac{1}{\sqrt{2}}\left( a_x - a_x^\dagger \right) = \sqrt{\frac{\gO}{2} } \left( b_x - b_x^\dagger \right)
~,
}
and the same for $y,p_y$\,. Consequently, the canonically deformed version of the anharmonic Hamiltonian \eqref{IsotropicOscillator:Hamiltonian} now becomes
\equ{
\label{QM_CanonicallyDeformedHamiltonian}
H= \frac{\textbf p^2}{2m}  + \frac{m\,\go^2_0 \,\textbf r^2}{2} 
+ \frac{m^N}{V^{N-1}}\, \frac{\gl}{2N}\, \textbf r^{2N}
\quad  \rightarrow \quad
H_{\gO}  = \frac{\gO\,\textbf p^2}{2m} + \frac{m\,\go^2_0 \,\textbf r^2}{2\, \gO} +\frac{m^N}{V^{N-1}}\, \frac{\gl}{2N}\,  \frac{ \textbf r^{2N} }{\gO^N}
~.
}
Notice also, that  charge conservation \eqref{ZeroModes:ChargeFixation} is compatible with this canonical  flow, in the sense that the angular momentum operator is not influenced by it,
%
$
 Q = x p_y - y p_x
 ~ \rightarrow ~
 Q_\gO =  Q
 ~.
 $


\subsection{Matrix elements of the su(1,1)}

\label{app:QM_Identities}

Here we show how to compute matrix elements, such as those invoked in the \textsc{vevs} 
\equ{
\label{GeneralizedCoherentVEVs_DEFINITION}
\braket{...} = \frac{\braket{\gz \vert ... \vert \gz} }{ \braket{\gz\vert \gz} }
~,
}
of Eq.\ \eqref{RadialCoherent:BarrierTerm} and \eqref{su11MatrixElement_potential}\,, in the radial coherent basis $\Ket{\gz}$ introduced in \eqref{RadialCoherentStateDEFINITION}. 
Take for example $K_-$\,:
\equ{
\begin{align}
\frac{\text{d}}{\text d \gz^*} \left[ \left( 1 - \vert \gz \vert^2 \right)^{-\frac{Q_0}{2} } \langle \gz\vert \gz' \rangle \right]
=&\,\,
\frac{\text{d}}{\text d \gz^*} \left[ \left( 1 - \vert \gz' \vert^2 \right)^{\frac{Q_0}{2}} \prescript{}{Q_0}{\langle 0\vert} \text{e}^{\gz^* K_-} \,\text{e}^{\gz' K_+} \vert 0 \rangle_{Q_0} \right]
=
\left( 1 - \vert \gz \vert^2 \right)^{-\frac{Q_0}{2}} \langle \gz \vert K_- \vert \gz' \rangle
~,
\end{align}
}
where $K_- \Ket{0}_{Q_0}=0$ is the vacuum state inside $\mathfrak H_{Q_0}$ defined by Eq.\ \eqref{RestrictedHilbertSpaceDEFINITION}\,.
At the same time, 
we can compute the derivative,
\equ{
\frac{\text{d}}{\text d \gz^*} \left[ \left( 1 - \vert \gz \vert^2 \right)^{-\frac{Q_0}{2} } \langle \gz \vert \gz' \rangle \right]
=
\frac{\left(1- \vert \gz'\vert^2 \right)^\frac{Q_0}{2} }{\left( 1 - \gz^* \gz' \right)^{Q_0+1}} \, Q_0 \, \gz'
~,
}
so that in total we find
\equ{
\Rightarrow \quad
\langle \gz \vert K_- \vert \gz'  \rangle = \frac{\left(1- \vert \gz'\vert^2 \right)^\frac{Q_0}{2} \left(1- \vert \gz\vert^2 \right)^\frac{Q_0}{2} }{\left( 1 - \gz^* \gz' \right)^{Q_0+1}} \, Q_0 \, \gz'
=
\frac{Q_0\, \gz'}{\left( 1 - \gz^* \gz' \right)} \, \langle \gz \vert \gz'  \rangle
~.
}
Along the same lines goes the matrix element of $K_+$\,.
Next, noting that 
\equ{
K_0 \left( K_+ \right)^n = n \left( K_+ \right)^n + \left( K_+ \right)^n K_0
~,
}
we use the series definition of $\text{e}^{\gz' K_+}$ to calculate 
\equ{
\begin{align}
\langle \gz  \vert K_0 \vert \gz' \rangle = 
\prescript{}{Q_0}{\langle 0 \vert} \text{e}^{\gz^* K_-} \,K_0 \,\text{e}^{\gz' K_+} \vert 0\rangle_{Q_0}
=
\langle \gz \vert \left( \frac{Q_0}{2} + \gz' \,K_+ \right) \vert \gz' \rangle
=
\frac{Q_0}{2} \, \frac{1 + \gz^* \gz' }{1- \gz^* \gz' } \,\langle \gz \vert \gz'\rangle
~,
\end{align}
}
where in the last equality we used the previously outlined $\langle \gz , K \vert K_+ \vert \gz', K \rangle$\,. 
The generalization to higher powers of $K_0,K_+,K_-$ is obvious.
In particular, to reach formula \eqref{su11MatrixElement_potential} it is needed to estimate the matrix-element of powers of the radial position operator $\textbf r^2$\,, e.g.\ calculating $\braket{\textbf r^4}$ we have
\begin{align}
\braket{\textbf r^4} =&\, \braket{(2K_0-K_+-K_-)^2}
\\
=&\,
\frac{Q_0}{2} \left\lbrace \left(Q_0+1\right) \left[ (\cosh^2\theta-1)(3+\cos2\phi)-4\cosh\theta\,\sqrt{\cosh^2\theta-1}\cos\phi \right] + 2\cosh\theta+ 2Q_0+4 \right\rbrace
~,
\nonumber
\\
=&\,
\braket{\textbf r^2}^2 +  \mathcal O (Q_0)
~.
\nonumber
\end{align}

\subsection{Thermal sums and loop integrals}
\label{sc:ThermalSums}

In this appendix we list the two basic types of loop sums encountered, when integrating-out the massive field to compute higher terms in the effective action \eqref{PhiN:EffectiveActionChiRESULT}\,.
The thermal sum over tree-level $a$-propagator $D_0(k)_a$ defined by \eqref{PhinN_aTreeLevelPropagator} 
in $d=D-1$ is resolved by
\begin{align}
\label{BasicLoop_MatsubaraSum_SCALING}
\sum_k D_0(k)_a
=&\,\,
V \int^{\vert\boldsymbol\gL\vert}\frac{\text{d}^d\textbf k}{(2\pi)^d}\, \sum_n  \left[\left(\frac{2\pi n}{\gb}\right)^2+\go^2_a(\textbf{k})\right]^{-1}
=
V \int^{\vert\boldsymbol\gL\vert}\frac{\text{d}^d\textbf k}{(2\pi)^d}\, \frac{\gb \coth \frac{\gb \go_a(\textbf{k})}{2} }{2 \go_a(\textbf{k})}
\nonumber
\\[2ex]
=&\,\,
\frac{\gb V}{2}\,
 \frac{\gG\left(\frac d2\right)}{(2\pi)^\frac{d}{2}}\,
 \frac{1}{d\sqrt{2\left(N-1\right)} }\,
 \frac{\vert\boldsymbol\gL\vert^d}{\left(\gl \rho_0^{N-1} \right)^\frac{1}{N+1} } 
 \,+\, \mathcal O \left( \frac{\vert\boldsymbol\gL\vert^{d+2}}{ \rho_0^\frac{3N-3}{N+1} } \right) 
 \, < \,
\mathcal O \left( \frac{1}{\rho_0^\frac{N-2}{N-1} } \right)
~.
\end{align}
%
The final order-estimation is done using the cut-off \eqref{TreeLevelMomentumCutOff_FINAL}\,.
Another basic thermal sum appearing in the  higher $a$-loops describes coupling to the source $j_\chi(x)$ of the form
\equ{
\label{LinearCurrentLoop_SCALING}
\int \text{d}^D x \, j_{\chi}(x) \, \sum_k \text{e}^{-ik\cdot x}\,  D_0(k)_a
\, < \,
\mathcal O \left( \left( \frac{1}{\rho_0}\right)^{\frac{2N-1}{2N+2}+ \frac{1}{2d(N+1)} } \right)
 ~,
}
where we can estimate the $\rho_0$-scaling of the position-space thermal propagator \cite{LANDSMAN1987141} as
\equ{
\label{SpaceTimePropagatorSCALING}
\left\vert \sum_k \text{e}^{-ik\cdot x}\,  D_0(k)_a \right\vert 
\leq
\frac{1}{\mu} \left \vert \int_0^{\vert \boldsymbol \gL \vert} \dd^d k \, \text{e}^{-ik\cdot x} \right\vert
\, \sim \,
\mathcal O \left( \frac{ \vert\boldsymbol \gL \vert^\frac{d-1}{2} }{\vert \textbf x \vert^\frac{d+1}{2}  }\, \frac{1}{\rho_0^\frac{N-1}{N+1} } \right)
\, < \,
\mathcal O \left( \left( \frac{1}{\rho_0}\right)^{\frac{2N-3}{2N+2}+ \frac{1}{2d(N+1)} } \right)
~.
}
%


\section{Compactification of large-charge theory}
\label{Appendix:Compactifications}

 \subsection{From flat space to the two-sphere 
 }
 \label{Appendix:SphereQuantization}
 
 
For the application of the derived formalism to sphere compactification 
we need to relate the flat action in $\mathbb R^{1,2}$ to the one on $\mathbb R_t \times S^2(R)$
 or after analytic continuation from $S^1(\beta) \times \mathbb R^2$ to $S^1(\beta) \times  S^2(R)$\,.
 For that we need the 
 solutions to Beltrami-Laplace eigenvalue problem,
\equ{
R^2 \Delta_{S^2_R}\,  Y^{m}_{l} = -l(l+1) Y^{m}_{l}
~,
}
with both $l,m \in \mathbb Z$ and $l\geq0$\,, while $m=-l,...,l$\,.
These are the so-called spherical harmonics,
 \begin{equation}
 Y^m_l(\Omega) = (-1)^m \sqrt{\frac{(2l+1)}{4\pi} \frac{(l-m)!}{(l+m)!} } \, P_l^m(\cos\theta) \, \text{e}^{im\phi}
 \end{equation}
 where we abbreviate $\Omega \equiv (\theta,\phi)$\,.
 Here, we use the Condon and Shortley phase convention, but this is irrelevant for the outcome.
 We record the properties of this orthonormal basis
 \equ{
  \label{Yharmonics:Orthogonality}
 \int_\Omega  \left(Y^m_l(\Omega)\right)^*\,  Y^{m'}_{l'}(\Omega) = \delta_{ll'}\, \delta_{mm'}
 \quad,\quad
\left(Y^m_l(\Omega)\right)^* = (-1)^m\,Y^{-m}_l(\Omega)
~,
 }
which in particular imply that
\equ{
\label{Yharmonics:ModifiedOrthogonality}
 \int_\Omega  Y^m_l(\Omega)\,  Y^{m'}_{l'}(\Omega) = (-1)^m\, \delta_{ll'}\, \delta_{m,-m'}
  \quad,\quad
%
\sum_{l,m}  Y^m_l(\Omega)\,  \left(Y^m_l(\Omega')\right)^* = \frac{1}{\sin\theta} \delta^{(2)}(\Omega-\Omega')
~.
}

\subsubsection*{The action on $S^2(R)$}

The Euclidean action of the free real scalar field $\psi(t,\Omega)$ 
on $ S^2(R)$ reads
\equ{
\label{ActionOnS2}
S[\psi] = \tfrac12 \int_0^\gb \text{d}\gt \int R^2 \,\text{d}\gO  \left[ \dot\psi^2 
+\psi \gD_{S^2_R} \psi
- m ^2 \, \psi^2\right]
~,
}
with the angular differential element of unit sphere $\text d \gO = \sin\vartheta \,\text{d}\vartheta \,\text{d}\varphi$\,. 
It turns out to be convenient to rescale our field operator as
\equ{
\psi(t,\Omega) = \frac{\phi(t,\Omega)}{R}
\quad \rightarrow \quad
[\phi] = -\frac12 ~,~ [\pi]=\frac12
~.
}
such that the rescaled field operator is expanded on $S^2(R)$ as
 \equ{
 \label{SphericalY:CanonicalField}
 \phi(t,\Omega) = 
\sum_{l,m} \frac{1}{\sqrt{2\omega} }\left(\text{e}^{i\pi \vert m \vert/2}\, \text{e}^{-i\omega t} \, Y^m_l(\Omega)\, a(l,m)
 +
\text{e}^{-i\pi \vert m \vert/2}\,  \text{e}^{i\omega t} \, \left(Y^m_l(\Omega)\right)^*\, a^\dagger(l,m)
 \right)
 ~,
 }
 where the ladder operators satisfy canonical (``flat'') commutation relations
 \equ{
 [ a(l,m)\,,\, a^\dagger(l',m')] = \delta_{ll'}\, \delta_{mm'}
 \quad , \quad
 [ a(l,m)\,,\, a(l',m')] = 0
 ~.
 }
The energy is evaluated at the discretized momentum points, i.e. $ \omega = \omega (p^2=\frac{l(l+1)}{2})$\,.
Starting from the action \eqref{ActionOnS2} 
with $\pi = 
\frac{\delta \mathcal L}{\delta \dot\phi} = \frac{ \sqrt{g} }{R^2}\, \dot \phi = \sin\vartheta\, \dot\phi$\,,
the Hamiltonian is then computed to
\equ{
H = \tfrac12\int_0^{2\pi} \text{d}\phi \int_0^\pi \text{d}\theta \sin\theta \left[ \frac{\pi^2}{\sin^2\theta} 
-  \phi \,\gD_{S^2_R}\, \phi
+ m^2 \, \phi^2\right]
~.
}
We can explicitly verify  that canonical quantization means,
\equ{
\left[\phi(t,\Omega)\,,\, \pi(t,\Omega') \right] = \, i \, 
\delta^{(2)}(\Omega - \Omega')
\quad\Leftrightarrow \quad
\left[ a(l,m) \,,\, a^\dagger(l',m') \right] =\,   \delta_{ll'}\, \delta_{mm'}
~,
}
so that 
plugging in expansion \eqref{SphericalY:CanonicalField} 
%
we compute for example the kinetic term,
\equ{
\int_0^{2\pi} \text{d}\phi \int_0^\pi \text{d}\theta \frac{\pi^2}{\sin\theta} 
=\sum_{l,m} \omega  \left\lbrace a^\dagger(l,m) a(l,m) -\tfrac12\left(a(l,-m)\, a(l,m) +  a^\dagger(l,-m)\, a^\dagger(l,m) \right) \right\rbrace
~,
}
where $\omega\equiv\omega(l,m)$\,. 
To resolve the angular integration formulae \eqref{Yharmonics:Orthogonality} and \eqref{Yharmonics:ModifiedOrthogonality} 
were used as well as the fact that $(-1)^m = (-1)^{\vert m \vert} $ and $\text{e}^{2\pi i \vert m \vert }=1$ for $m \in \mathbb Z$\,.
We recognize that this expression has the same formal structure as its flat space counter-part, which allows us to recycle the developed formalism in Section \ref{sc:QuantumFluctuations}.

\subsection{A  toy model on $S^1(R)$}
\label{appedix:ToyModelS1}


To demonstrate the  Casimir effect for the $O(2)$ model at large charge, we consider a toroidal compactification.
In particular, for computational simplicity we focus on
$3+1$ space-time dimensions and compactify one spatial direction (say  $z$) on a circle of circumference $2\pi R$\,, i.e.\ our theory lives in $S^1(\gb) \times \mathbb R^2 \times S^1(R)$\,.
This results in periodic \textsc{bc} along this direction
and vanishing along the other two encoded as $\boldsymbol r = (x,y)$\,:
\equ{
\label{PeriodicIdentifications_BC}
z \sim z + 2\pi R  \quad \text{ with field \textsc{bc}}
\\[1ex]
\phi(\gt +\gb , \textbf r) = \phi(\gt, \textbf r)
\quad, \quad
\phi(\gt , \boldsymbol r,z+2\pi R) = \phi(\gt, \boldsymbol r,z)
\quad, \quad
\phi(\gt , \boldsymbol r \rightarrow \infty,z) = 0
~.
\nonumber
}
The momentum  along the compactified direction gets discretized accordingly, $k_z= \tfrac{n}{R}$,
such that the spherical cut-off \eqref{TreeLevelMomentumCutOff_FINAL} now becomes
\equ{
\label{CasimirOnS1_SphericalCutOff}
n \leq R\, \vert \boldsymbol \gL \vert 
\quad \text{and} \quad
k_x^2+k_y^2 \, \leq \, \vert \boldsymbol \gL \vert^2 - \frac{n^2}{R^2}
~.
}
Using then the leading Goldstone dispersion relation from \eqref{PhiN:EffectiveActionChiRESULT}, 
\equ{
\go_\chi({\textbf k}) = \sqrt{\frac{N-1}{N+1} } \vert \textbf k \vert
~,
}
the Casimir energy per unit area along the two uncompactified directions becomes
\equ{
\label{CasimirOnS1_FinalResult_SphericalSym}
\frac{\mathcal F^{(p)}_V }{L^2} = 
- \frac{1}{24\pi}\, \sqrt{\frac{N-1}{N+1} }\, \frac{\vert\boldsymbol\gL\vert^2}{R } 
~<~
\mathcal O \left(\rho_0^\frac{2}{3(N+1)} \right)
~,
}
where we may think of the total volume as $V=L^2 \times (2\pi R)$ with the understanding that eventually $L\rightarrow \infty$\,.
Concerning the $\rho_0$-scaling of the cut-off $\boldsymbol\gL$ it is admissible to implement \eqref{TreeLevelMomentumCutOff_FINAL}\,, since we have chosen to work with a spherically compatible cut-off as the one introduced in  \eqref{CasimirOnS1_SphericalCutOff}\,.

Higher dimensional (at least toroidal) compactifications will result in more involved lattice sums.
Still, what we recognize from the  simple toy model on $S^1(R)$, is that because of the fundamental momentum cut-off \eqref{TreeLevelMomentumCutOff_FINAL} any lattice sum will be finite and shall contribute at a given $\rho_0$-order.

\section{Free field theory at fixed charge}
\label{appendix:FreeTheory}

To complete the treatment of scalar models at fixed charge, 
we deal here with the $O(2)$ action of a free complex scalar field. 
Note however, that this is essentially different from the \textsc{bec} procedure and the coherent state we constructed in Section \ref{ssc:CoherentCOnstructionOfZeroModeVacuum} for the interacting case.
This compact treatment of the free theory using the holomorphic language  \cite{zinn2010path} is done in a way that might shed some light onto future extensions of fixed charge constructions, especially in the direction of the purely fermionic scenario. 


First, we consider the relevant quantum mechanical system, namely the harmonic isotropic oscillator in 2D,
described by Hamiltonian \eqref{IsotropicOscillator:Hamiltonian} for $\gl=0$\,.
In the circular basis \eqref{QM:CircularOperatorBasis} for some arbitrary frequency $\go$,
it simply reads (disregarding the normal ordering constant)
\equ{
\label{IsotropicHarmonicHamiltonian}
H = \go \left( a_L^\dagger a_L + a_R^\dagger a_R \right)
~.
}
As we have seen in Section \ref{ssc:CoherentCOnstructionOfZeroModeVacuum}\,, in the appropriate basis $\lbrace \Ket{n_L,n_R} \rbrace$
fixing the angular momentum to $Q_0$ as in Eq.\ \eqref{ZeroModes:ChargeFixation_OperatorIdentity} implies the restricted Hilbert space \eqref{RestrictedHilbertSpaceDEFINITION}\,.
Hence, it is straight-forward to compute the fixed charge partition sum of the isotropic oscillator inside $\mathfrak h_{Q_0}$ to
\equ{
Z_{Q_0}(\gb) \,:= \sum_{\substack{n_L, n_R \\ \vert n_L-n_R\vert =Q_0}} \text{e}^{-\gb\go(n_L+n_R)}
= 
\frac{2\, \text{e}^{-\gb\go Q_0} - \gd_{Q_0,0}}{1-\text{e}^{-2\gb\go}}~.
}

To compute the time-evolution operator we now pass to the holomorphic formalism.
This is done mainly by promoting the canonical ladder operators to holomorphic variables.
In this formalism we first define the kernel of time evolution operator at level $Q_z$ (the projection of $Q_0$ along the $z$-axis) by
\equ{
\begin{align}
\mathcal U_{Q_z} (\bar \ga^\gb,\ga^0) 
:= &\,\,
\Bra{\bar \ga^\gb} \gd(Q-Q_0) \,\text{e}^{-\gb H} \Ket{a^0}
\equiv 
 \int \frac{\text{d}\theta}{2\pi} \text{e}^{-i\theta |Q_z|}~ \mathcal U_L^\theta(\bar \ga^\gb_L,\ga^0_L)~ \mathcal U_R^\theta(\bar \ga^\gb_R,\ga^0_R)~, 
\end{align}
}
where $a_i\equiv a_i(\gt)$ and $\bar a_i \equiv \bar a(\gt)$ for $i=L,R$ are thought of as independent variables and the two $\theta$-twisted kernels are given by
\equ{
\begin{align}
\mathcal U_i^\theta(\bar \ga^\beta_i,\ga^0_i) =
\exp \left\lbrace \bar a^\gb_{L}\, a^0_{L}~ \text{e}^{-\beta \go \pm i\theta} \right\rbrace\,
\quad \text{ with }~~ i = L,R
\quad\text{and}\quad + \text{ for } L ~,~ - \text{ for } R
~.
\end{align}
}
$\bar \ga^\gb \equiv (\bar \ga_L^\gb , \bar \ga_R^\gb)$ and $\ga^0 \equiv (\ga_L^0 , \ga_R^0)$ denote collectively the 
temporal 
\textsc{bc} for left- and right-movers, i.e.\ $a_i(0) = \ga_i^0$ and $\bar a_i(\gb) = \ga_i^\gb$\,.
%
Using Hamiltonian \eqref{IsotropicHarmonicHamiltonian} we find for the time-evolution operator
\equ{
\label{hol:TimeEvolution_J0_RESULT}
\mathcal U_{Q_z}(\bar \ga^\gb, \gb; \ga^0,0) 
= \text e^{-i\phi |Q_z|}\, I_{Q_z}(2 \sqrt{\mathcal A_L \mathcal A_R})~,
}
where $I_m(x)$ denotes the modified Bessel function of the first kind and the phase shift is dictated by 
\equ{\label{
hol:AngleShift_TimeEvolutionJ0}
\tan \phi = i\,\frac{\mathcal A_L - \mathcal A_R}{\mathcal A_L + \mathcal A_R}~, \quad -\frac \pi 2 < \phi < \frac \pi 2~,
}
in terms of left-/right-moving amplitudes,
\equ{
\mathcal A_L = \bar \ga_L^\gb\, \ga_L^0 \, \text e^{-\gb\go}~, \quad A_R = \bar \ga_R^\gb \,\ga_R^0 \, \text e^{-\gb\go}~.
}
Now, we are in a position to introduce the full time-evolution operator at level $Q_0=\vert Q_z \vert$ as the sum of the two $z$-projected ones,
\equ{
\mathcal U_{Q_0} ( \ga^\gb ,   \ga^0 )
=
\mathcal U_{Q_z} (\ga^\gb ,\ga^0 )+ \mathcal U_{-Q_z} (\ga^\gb ,\ga^0 )
=
\left( 2 - \gd_{Q_0,0} \right) \cos \left( \phi\, Q_0 \right)\, I_{Q_0}(2 \sqrt{\mathcal A_L \mathcal A_R})
~.
}


The immediate generalization of the preceding quantum mechanical treatment to $(d+1)$-dimensional quantum field theory takes \cite{faddeev1980gauge} the form 
\equ{\label{FreeQFT_PartitionSumFixedQ0_Unresolved}
\begin{align}
&
Z_{Q_0}(\gb) = \Tr \left\lbrace\gd \left(Q-Q_0 \right)\text{e}^{-\gb  H}\right\rbrace = 
 \int_{0}^{2\pi} \frac{\text d\theta}{2\pi}\, \text{e}^{-i\theta Q_0}\, \Tr \left(\text{e}^{i\theta Q}\text{e}^{-\gb  H}\right)
\\[1ex]
&=\,
 \int_{0}^{2\pi} \frac{\text d\theta}{2\pi}\, \text{e}^{-i\theta Q_0}\, 
 \prod_{\textbf k} \frac{1}{1-\text e^{-\gb\go_\textbf{k}+i\theta}}\, \frac{1}{1-\text e^{-\gb\go_\textbf{k}-i\theta}} \nonumber
= 
\sum_\textbf{k} \frac{\text{e}^{-\gb\go_{\textbf k}Q_0}}{1-\text{e}^{-2\gb\go_{\textbf k}}} \prod_{\textbf q \neq \textbf k} \frac{1}{1-\text e^{-\gb(\go_\textbf{q}-\go_\textbf{k})}} \frac{1}{1-\text e^{-\gb(\go_\textbf{q}+\go_\textbf{k})}} ~,
\end{align}
}
with the mass-shell condition $\go_\textbf{k} = \sqrt{ M^2 + \textbf k^2}$ and the charge operator $Q$ given in \eqref{ZeroModes:ChargeFixation_OperatorIdentity}\,.

\pagebreak

{\small
\providecommand{\href}[2]{#2}\begingroup\raggedright\endgroup
} 


\begin{thebibliography}{10}

\bibitem{Alvarez-Gaume:2016vff}
L.~Alvarez-Gaume, O.~Loukas, D.~Orlando, and S.~Reffert ``{Compensating strong
  coupling with large charge}''
\href{http://www.arXiv.org/abs/1610.04495}{[{\tt arXiv:1610.04495}]}.

\bibitem{Hellerman:2015nra}
S.~Hellerman, D.~Orlando, S.~Reffert, and M.~Watanabe ``{On the CFT Operator
  Spectrum at Large Global Charge}'' {\em JHEP} {\bf 12} (2015) 071
\href{http://www.arXiv.org/abs/1505.01537}{[{\tt arXiv:1505.01537}]}.

\bibitem{Nicolis:2011pv}
A.~Nicolis and F.~Piazza ``{Spontaneous Symmetry Probing}'' {\em JHEP} {\bf 06}
  (2012) 025
\href{http://www.arXiv.org/abs/1112.5174}{[{\tt arXiv:1112.5174}]}.

\bibitem{Watanabe:2013uya}
H.~Watanabe, T.~Brauner, and H.~Murayama ``{Massive Nambu-Goldstone Bosons}''
  {\em Phys. Rev. Lett.} {\bf 111} (2013) no.~2, 021601
\href{http://www.arXiv.org/abs/1303.1527}{[{\tt arXiv:1303.1527}]}.

\bibitem{MurayamaLecture}
{H. Murayama} ``{Quantum Field Theory II (Bose Systems)}.'' Lecture notes 2005.
\newblock
  \href{http://hitoshi.berkeley.edu/221B/bosons.pdf}{http://hitoshi.berkeley.edu/221B/bosons.pdf}.

\bibitem{Monin:2016jmo}
A.~Monin, D.~Pirtskhalava, R.~Rattazzi, and F.~K. Seibold ``{Semiclassics,
  Goldstone Bosons and CFT data}''
\href{http://www.arXiv.org/abs/1611.02912}{[{\tt arXiv:1611.02912}]}.

\bibitem{inomata1992path}
A.~Inomata, H.~Kuratsuji, and C.~Gerry {\em Path Integrals and Coherent States
  of SU(2) and SU(1,1)}.
\newblock World Scientific 1992.

\bibitem{klauder1985applications}
J.~Klauder and B.-S. Skagerstam ``{Applications in physics and mathematical
  physics}'' {\em World Scientific, Singapore} (1985).

\bibitem{MLODINOW1980314}
L.~Mlodinow and N.~Papanicolaou ``{SO(2, 1) algebra and the large N expansion
  in quantum mechanics}'' {\em Annals of Physics} {\bf 128} (1980) no.~2, 314
  -- 334.

\bibitem{MLODINOW1982387}
L.~D. Mlodinow ``{Large-N expansions work: A semi-classical perturbation theory
  for quantum mechanics}'' {\em Progress in Particle and Nuclear Physics} {\bf
  8} (1982) 387 -- 399.

\bibitem{arecchi1972atomic}
F.~Arecchi, E.~Courtens, R.~Gilmore, and H.~Thomas ``{Atomic coherent states in
  quantum optics}'' {\em Physical Review A} {\bf 6} (1972) no.~6, 2211.

\bibitem{perelomov1972}
A.~M. Perelomov ``{Coherent states for arbitrary Lie group}'' {\em Comm. Math.
  Phys.} {\bf 26} (1972) no.~3, 222--236.

\bibitem{PhysRevA.38.191}
C.~C. Gerry and J.~Kiefer ``{Radial coherent states for central potentials: The
  isotropic harmonic oscillator}'' {\em Phys. Rev. A} {\bf 38} (Jul, 1988)
  191--196.

\bibitem{gerry1983large}
C.~C. Gerry, J.~B. Togeas, and S.~Silverman ``{Large-N phase-integral
  approximation for SU(1, 1) coherent states}'' {\em Physical Review D} {\bf
  28} (1983) no.~8, 1939.

\bibitem{kapusta2006finite}
J.~I. Kapusta and C.~Gale {\em {Finite-temperature field theory: Principles and
  applications}}.
\newblock Cambridge University Press 2006.

\bibitem{Argyres:2009em}
E.~N. Argyres, C.~G. Papadopoulos, M.~T.~M. van Kessel, and R.~H.~P. Kleiss
  ``{Path Integrals in Polar Field Variables in QFT}'' {\em Eur. Phys. J.} {\bf
  C61} (2009) 495--518
\href{http://www.arXiv.org/abs/0901.0815}{[{\tt arXiv:0901.0815}]}.

\bibitem{Nicolis:2013sga}
A.~Nicolis, R.~Penco, F.~Piazza, and R.~A. Rosen ``{More on gapped Goldstones
  at finite density: More gapped Goldstones}'' {\em JHEP} {\bf 11} (2013) 055
\href{http://www.arXiv.org/abs/1306.1240}{[{\tt arXiv:1306.1240}]}.

\bibitem{Alday:2016jfr}
L.~F. Alday ``{Solving CFTs with Weakly Broken Higher Spin Symmetry}''
\href{http://www.arXiv.org/abs/1612.00696}{[{\tt arXiv:1612.00696}]}.

\bibitem{Xiao:1198089}
M.-w. Xiao ``{Theory of transformation for the diagonalization of quadratic
  Hamiltonians}'' Tech. Rep. arXiv:0908.0787 Aug, 2009.

\bibitem{PhysRevD.29.643}
C.-S. Hsue and J.~L. Chern ``Two-step approach to one-dimensional anharmonic
  oscillators'' {\em Phys. Rev. D} {\bf 29} (Feb, 1984) 643--647.

\bibitem{LANDSMAN1987141}
N.~Landsman and C.~van Weert ``{Real- and imaginary-time field theory at finite
  temperature and density}'' {\em Physics Reports} {\bf 145} (1987) no.~3, 141
  -- 249.

\bibitem{zinn2010path}
J.~Zinn-Justin {\em Path Integrals in Quantum Mechanics}.
\newblock Oxford Graduate Texts. OUP Oxford 2010.

\bibitem{faddeev1980gauge}
L.~D. Faddeev and A.~Slavnov ``{Gauge fields: Introduction to Quantum
  Theory}''.

\end{thebibliography}
\end{document}